\documentclass[useAMS,usenatbib]{mn2e}

\usepackage{times}
\usepackage{graphics,epsfig}
\usepackage{graphicx}
\usepackage{amsmath}
\usepackage{amssymb}
\usepackage{rotating}
\usepackage{ulem}
\usepackage{float}

\newcommand{\ha}{\ensuremath{{\rm H}\alpha}}
\newcommand{\ms}{\ensuremath{\rm M_{\odot}}}

\newcommand{\um}{\ensuremath{\rm \mu m}}
\newcommand{\sgas}{$\Sigma_{\rm gas}$}
\newcommand{\shi}{$\Sigma_{\rm gas,atomic}$}
\newcommand{\ssfr}{$\Sigma_{\rm SFR}$}

\newcommand{\sfh}{\ensuremath{\Sigma_{\rm SFR}^{H\alpha}}}
\newcommand{\shtwo}{\ensuremath{\Sigma_{\rm H_2}}}

\title[Atomic gas and SFR in irregulars]{The relation between atomic gas and star formation rate densities in faint irregular galaxies}

\author[Roychowdhury et al.]{Sambit Roychowdhury,$^{1}$\thanks{E-mail: sambit@mpa-garching.mpg.de (SR); chengalu@ncra.tifr.res.in (JNC); skai@sao.ru (SSK); ikar@sao.ru (IDK)} Jayaram N. Chengalur,$^{2\star}$ Serafim S. Kaisin$^{3\star}$ 
\newauthor and Igor D. Karachentsev$^{3\star}$\\
       \\ 
       $^{1}$Max-Planck-Institut f\"{u}r Astrophysik, Karl-Schwarzschild-Str. 1, 85748 Garching, Germany\\
       $^{2}$NCRA-TIFR, Post Bag 3, Ganeshkhind, Pune 411 007, India\\
       $^{3}$Special Astrophysical Observatory, Russian Academy of Sciences, N. Arkhyz, KChR 369167, Russia}

\begin{document}
\date{}

\pagerange{\pageref{firstpage}--\pageref{lastpage}} \pubyear{}

\maketitle

\label{firstpage}

\begin{abstract}

  We use data for faint ($\rm{M_B~>~-14.5}$) dwarf irregular galaxies drawn from the FIGGS survey to study the correlation between the  atomic gas density (\shi) and star formation rate (\ssfr) in the galaxies. The estimated gas phase metallicity of our sample galaxies is $Z \sim 0.1$~Z$_{\odot}$. Understanding star formation in such molecule poor gas is of particular importance since it is likely to be of direct relevance to  simulations of early galaxy formation. For about 20\% (9/43) of our sample galaxies, we find that the HI distribution is significantly disturbed, with little correspondence between the optical and HI distributions. We exclude these galaxies from the comparison. We also exclude galaxies with very low star formation rates, for which stochastic effects make it difficult to estimate the true star formation rates. For the remaining galaxies we compute the \shi\ and \ssfr\ averaged over  the entire star forming disk of the galaxy. For these galaxies we find a nearly linear relation between the star formation rate and the atomic gas surface densities, viz.
$\rm{\log \Sigma_{SFR} = 0.91^{+0.23}_{-0.25} \log \Sigma_{gas,atomic} - 3.84^{+0.15}_{-0.19}}$. 
The corresponding gas consumption timescale is $\sim$ 10 Gyr, i.e. significantly smaller than the $\sim 100$~Gyr estimated for the outer regions of spiral galaxies. We also estimate the gas consumption timescale computed using the global gas content and the global star formation rate for all galaxies with a reliable measurement of the star formation rate, regardless of whether the HI distribution is disturbed or not. The mean gas consumption timescale computed using this entire gas reservoir is $\sim 18$~Gyr, i.e. still significantly smaller than that estimated for the outer parts of spirals. The gas consumption timescale for dwarfs is intermediate between the values of $\sim 100$~Gyr and $\sim 2$~Gyr estimated for the outer molecule poor and inner molecule rich regions of spiral disks.

\end{abstract}


\section{Introduction}
\label{sec:int}

The rate at which gas is converted into stars in galaxies is an important input into understanding galaxy formation and evolution. However, despite decades of studies it remains poorly understood, mainly because, the exact processes that govern this transition are complex \citep[see e.g.][]{McK07} and difficult to capture in their entirety in analytical or numerical models. Most models hence use empirical star formation ``recipes'' to model the formation of stars \citep[e.g. see][]{sch08,gov10,hop13,fu13}. These recipes are generally based  on scaling relations between gas density and star formation rate that have been established via observations of nearby galaxies. One of the most commonly used parameterizations of this type is the   Kennicutt-Schmidt relation \citep{s59,ken98} which relates the surface densities of gas (\sgas) and star formation rate surface density (\ssfr) through a power law. The power law index found by \citet{ken98} (i.e. $N \sim 1.4$) was based on a sample of spiral and star bursting galaxies. The variation of this relationship with environment remains an area of active research. There have also been a number of recent studies of the correlation between the molecular gas surface density \shtwo\ and \ssfr, but there is  still no consensus regarding the values of the parameters of the power law used to characterise this correlation. For e.g. \citet{ler13} find the coefficient of the power law to be $N \sim 1 \pm 0.15$, while \citet{mom13} find that it could be as steep as $N \sim 1.8$.  Regarding the correlation between the atomic gas surface density \shi\ and the \ssfr\ the situation appears still more complex. In the central parts of spiral galaxies, the HI gas density appears to be uncorrelated to \ssfr\, while in the outer, molecule poor regions, the two appear to be significantly correlated \citep{big10}. Understanding star formation in molecule poor gas is of particular importance since it is likely to be of direct relevance in cosmological simulations of early galaxy formation. 

This paper is part of an ongoing study regarding star formation in molecule poor gas in nearby, very faint dwarf galaxies. Our studies are based on samples drawn from  the Faint Irregular Galaxy GMRT Survey \citep[FIGGS][]{beg08}. In \citet{roy09} it was shown that \shi\ and \ssfr\ are correlated on sub-kpc scales, and that in general the power law index is steeper than the value of 1.4 found by \citet{ken98}. In \citet{roy11} we showed (again on sub kpc scales) that all regions with \shi\ $\gtrsim$~ 10 \ms${\rm pc}^{-2}$ have some associated star formation, but that the fraction of gas with associated star formation decreases steadily as \shi\ decreases. Since star formation is likely to proceed via the formation of molecular gas, this means that the molecular fraction is significant for all gas with \shi\ $\gtrsim$~ 10 \ms${\rm pc}^{-2}$.  The number of FIGGS galaxies with FUV and \ha\ observations has increased significantly since the work of \citet{roy11}. {\it Spitzer} observations of dust emission are also available for several of these galaxies, which allows us to correct for dust extinction using the recent ``composite'' star formation calibrations. In this paper we use the new observational data, and the new calibrations to study the relationship between \shi\ and \ssfr\  averaged over the entire star forming disk, as well as computed using the total gas content and the total star formation rate.  We also compare our results with those obtained from studies of nearby spiral galaxies. The conditions in ISM in dwarf galaxies is expected to be similar to that of the outskirts of spiral disks, in that both are molecule poor. However, the faint dwarfs that we study here have somewhat lower metallicity (viz. Z~$\sim 0.1$~Z$_{\odot}$) than what is typical in the outskirts of spirals \citep[viz. Z~$\sim 0.4-0.6$~Z$_{\odot}$,][]{car07,gen14}. Further,  as compared to the outskirts of spirals, the gas in our sample galaxies experiences less rotational shear.
 
\section {Sample and Methodology}
\label{sec:samp}

\begin{table*}
\caption{Sample parameters}
\label{tab:samp}
\begin{tabular}{|lccccccccc|}
\hline
Galaxy&$\alpha$ (J2000)&$\delta$ (J2000)&M$_{\rm B}$&D&D${\rm{_{Ho}}}$&b/a&Z/Z$_{\odot}$&HI beam&\ha\ resolution\\
      &($^h~^m~^s$)&($^\circ$~$^\prime$~$^{\prime\prime}$)&&(Mpc)&($^\prime$)&&&$^{\prime\prime}~\times~^{\prime\prime}$&$^{\prime\prime}$\\
\hline
And IV           &00 42 32.30&$+$40 34 19&$-$8.39~&6.3~&1.00&0.77&0.06~~~~&13.9$\times$12.7&\\
DDO 226          &00 43 03.80&$-$22 15 01&$-$14.17&4.9~&2.24&0.36&0.12~~~~&16.8$\times$16.8&\\
DDO 6            &00 49 49.30&$-$21 00 58&$-$12.5~&3.34&2.29&0.41&0.07~~~~&26.5$\times$20.6&\\
UGC 685          &01 07 22.44&$+$16 41 04&$-$14.31&4.5~&2.40&0.71&0.20$^*$&16.8$\times$16.0&1.9\\
KK 14            &01 44 42.80&$+$27 17 19&$-$12.13&7.2~&0.79&0.37&0.06~~~~&13.4$\times$10.0&1.4\\
KK 41            &04 25 20.00&$+$72 48 30&$-$14.06&3.9~&3.72&0.57&0.11~~~~&25.8$\times$17.1&2.0\\
UGCA 92          &04 32 04.90&$+$63 36 49&$-$15.65&3.01&2.00&0.50&0.19~~~~&30.2$\times$24.2&\\
KKH 34           &05 59 40.40&$+$73 25 40&$-$12.30&4.6~&0.93&0.56&0.06~~~~&20.0$\times$16.0&\\
UGC 3755         &07 13 51.60&$+$10 31 19&$-$14.90&6.96&1.86&0.59&0.15~~~~&11.8$\times$11.1&1.5\\
DDO 43           &07 28 17.20&$+$40 46 13&$-$14.75&7.8~&1.41&0.69&0.14~~~~&11.7$\times$09.5&\\
KK 65            &07 42 31.98&$+$16 33 40&$-$14.29&7.62&1.12&0.56&0.12~~~~&11.5$\times$10.3&2.1\\
UGC 4115         &07 57 01.80&$+$14 23 27&$-$14.27&7.5~&1.91&0.56&0.12~~~~&11.0$\times$11.0&\\
KDG 52           &08 23 56.00&$+$71 01 46&$-$11.49&3.55&1.26&0.92&0.05~~~~&24.2$\times$21.5&\\
UGC 4459         &08 34 07.20&$+$66 10 54&$-$13.37&3.56&2.00&0.87&0.13$^*$&24.9$\times$21.1&1.5\\
UGC 5186         &09 42 58.66&$+$33 15 56&$-$12.98&6.9~&1.38&0.23&0.08~~~~&12.1$\times$10.5&1.9\\
UGC 5209         &09 45 04.20&$+$32 14 18&$-$13.15&6.7~&0.83&0.96&0.08~~~~&12.8$\times$10.7&\\
UGC 5456         &10 07 19.70&$+$10 21 44&$-$15.08&5.6~&1.62&0.50&0.16~~~~&14.7$\times$14.7&\\
UGC 6145         &11 05 35.00&$-$01 51 49&$-$13.14&7.4~&1.38&0.56&0.08~~~~&11.1$\times$11.1&\\
UGC 6456         &11 27 59.90&$+$78 59 39&$-$14.03&4.3~&1.48&0.53&0.10$^{\dagger}$&19.4$\times$17.0&2.5\\
UGC 6541         &11 33 29.10&$+$49 14 17&$-$13.71&3.9~&1.74&0.57&0.13$^*$&22.7$\times$21.4&\\
NGC 3741         &11 36 06.40&$+$45 17 07&$-$13.13&3.0~&1.48&0.55&0.09$^*$&28.2$\times$27.0&\\
DDO 99           &11 50 53.00&$+$38 52 50&$-$13.52&2.6~&4.27&0.37&0.10~~~~&31.7$\times$31.7&\\
E321$-$014       &12 13 49.60&$-$38 13 53&$-$12.70&3.2~&1.41&0.43&0.07~~~~&29.9$\times$21.6&\\
UGC 7242         &12 14 07.40&$+$66 05 32&$-$14.06&5.4~&1.23&0.42&0.11~~~~&15.3$\times$15.3&\\
CGCG 269$-$049   &12 15 46.63&$+$52 23 14&$-$13.25&4.9~&1.05&0.30&0.05$^*$&16.8$\times$16.8&2.0\\
UGC 7298         &12 16 30.10&$+$52 13 39&$-$12.27&4.21&0.85&0.55&0.06~~~~&19.6$\times$19.6&\\
KK 144           &12 25 29.15&$+$28 28 57&$-$12.59&6.3~&1.12&0.44&0.07~~~~&16.0$\times$10.3&1.6\\
DDO 125          &12 27 40.90&$+$43 29 44&$-$14.16&2.5~&3.89&0.56&0.12~~~~&34.8$\times$30.0&2.0\\
UGC 7605         &12 28 38.75&$+$35 43 03&$-$13.53&4.43&1.48&0.73&0.10~~~~&22.3$\times$14.8&2.0\\
GR8              &12 58 40.40&$+$14 13 03&$-$12.11&2.1~&1.66&0.91&0.09$^*$&39.3$\times$39.3&\\
UGC 8215         &13 08 03.60&$+$46 49 41&$-$12.26&4.5~&0.85&0.70&0.06~~~~&19.2$\times$17.9&\\
DDO 167          &13 13 22.80&$+$46 19 11&$-$12.70&4.2~&1.10&0.55&0.07~~~~&20.7$\times$18.8&\\
KK 200           &13 24 36.00&$-$30 58 20&$-$11.96&4.6~&1.10&0.62&0.06~~~~&17.9$\times$17.9&\\
E444$-$78        &13 36 30.80&$-$29 14 11&$-$13.3~&5.25&1.58&0.42&0.09~~~~&15.7$\times$15.7&\\
UGC 8638         &13 39 19.40&$+$24 46 32&$-$13.68&4.27&1.66&0.67&0.10~~~~&19.3$\times$19.3&2.0\\
DDO 181          &13 39 53.82&$+$40 44 21&$-$13.03&3.1~&2.40&0.57&0.14$^*$&29.8$\times$25.2&1.4\\
DDO 183          &13 50 51.10&$+$38 01 16&$-$13.17&3.24&2.40&0.32&0.09~~~~&27.4$\times$24.5&\\
UGC 8833         &13 54 48.70&$+$35 50 15&$-$12.42&3.2~&1.17&0.89&0.07~~~~&26.4$\times$25.2&\\
KK 230           &14 07 10.70&$+$35 03 37&$-$9.55~&1.9~&0.76&0.83&0.03~~~~&43.4$\times$43.4&\\
DDO 187          &14 15 56.50&$+$23 03 19&$-$12.51&2.5~&1.70&0.76&0.11$^*$&33.0$\times$33.0&\\
KK 246           &20 03 57.40&$-$31 40 54&$-$13.69&7.83&0.91&0.42&0.10~~~~&10.5$\times$10.5&\\
UGCA 438         &23 26 27.50&$-$32 23 26&$-$12.94&2.2~&2.14&0.80&0.08~~~~&37.5$\times$37.5&\\
KKH 98           &23 45 34.02&$+$38 43 04&$-$10.78&2.5~&1.05&0.55&0.04~~~~&34.5$\times$31.6&2.7\\
\hline
\end{tabular}
\begin{flushleft}
$^*$: Based on \citet{mar10},\\
$^{\dagger}$: based on \citet{mou06}.
\end{flushleft}
\end{table*}

The galaxy sample that we use in this paper are summarised in Table~\ref{tab:samp}. The columns in the table are: 
column (1) the name of the galaxy, 
columns (2) and (3) the coordinates in the J2000 system, 
column (4) the absoulte B-band mahnitude form \citet{beg08},
column (5) the distance in Mpc, 
column (6) the B-band diameter at 26.5 magnitude arcsecond$^{\rm -2}$ (i.e. the Holmberg diameter), 
column (7) the measured apparent axial ratio in the B-band.
The values for columns (5), (6) and (7) are from \citet{kar13}.
Column (8) gives the estimated gas phase metalicity (see Section~\ref{ssec:estimates} for details).
As mentioned above, the galaxies are drawn from the FIGGS sample -- the current subsample includes all galaxies in that sample with available FUV and/or \ha\ data. The FUV data are taken from the public {\it GALEX} archive. The \ha\ data are either from our observations using the 6m BTA telescope in Russia \citep{kar07,kai08,kar10,kai11}, or drawn from the literature if we did not have 6m BTA \ha\ map for a galaxy.
In Table~\ref{tab:samp} column (9) gives the FWHM of the HI beam for each galaxy in arcseconds. The choice of the beam size to make the neutral hydrogen maps is discussed in the following section.
Column (10) gives the resolution of the 6m BTA telescope \ha\ images.
{\it GALEX} FUV images have a resolution of $\sim$4.5 arcseconds.

\subsection{ Estimates of the Gas and Star Formation Rate density}
\label{ssec:estimates}

The aim of our study is to study the relationship between the surface densities of gas (\sgas) and star formation (\ssfr) for the sample galaxies. These surface densities are measured as the average over the `star-forming disk' of each galaxy, which we define as the Holmberg ellipse. We restrict the calculated values to the `star-forming disk' in order to make a comparison between the current star formation and gas present in the region where star formation is occurring. The assumption is hence that the remaining gas is not involved in the current star formation. We relax this assumption in Sec.~\ref{sec:totrel} and compute the relation between the total gas content and the total star formation rate. Figure~\ref{fig:ov} shows overlays of the HI, FUV and \ha\ emission along with the Holmberg ellipse for galaxies in the sample. As can be seen, the \ha\ emission is generally centrally concentrated and overlaps with the regions with the highest HI column density. For most galaxies, the FUV emission is also restricted to within  the Holmberg ellipse. This provides further justification for assuming that the gas within the Holmberg ellipse is most related to the ongoing star formation. For a handful of galaxies some very low level FUV emission lies outside the Holmberg ellipse. For these galaxies we increase the size of the axes for the `star-forming disk' by 10\%, (consistent with the estimate of maximum error on the Holmberg ellipse parameter measurements). The resultant ellipse now contains all the FUV emission. These galaxies are marked in Table~\ref{tab:flux} and in the overlays (see online version).

Total intensity HI maps were used to determine the average column density within the `star-forming disk'. In keeping with our previous work \citep{roy09,roy11} we made HI column density maps of all our sample galaxies at the uniform sub-kpc linear resolution of $\sim$400 pc. This was the best achievable resolution while ensuring that significant amount of extended low level emission is not missed due to the decreasing signal-to-noise with increasing resolution for all of our sample galaxies. The FWHM of our HI column density maps (listed in Table~\ref{tab:samp}) are comparable to those used for many previous works studying disk-averaged Schmidt type relations including \citet{ken98,wyd09}. The atomic gas density is corrected for the presence of helium by multiplying by a factor of 1.34. No correction is made for the presence of molecular gas (but see the discussion in Sec.~\ref{ssec:dis}). The face on surface densities are computed by
correcting for the inclination of the disks, assuming that they are oblate spheroids \citep[a valid assumption considering the luminosity range of the dwarf irregular galaxies in our sample, e.g. see][]{roy13}. We multiply the measured surface density by the cosine of the inclination angle in order to estimate the face on column density. The details of how the {\it GALEX} FUV and BTA \ha\ observations are converted to fluxes are given in \citet{roy09,roy11}. Briefly, foreground stars and background galaxies in the FUV and \ha\ images are masked. When calculating \ssfr, the flux of either tracer is averaged over the non-masked area for that tracer. For fluxes taken from the literature, surface densities are obtained by diving by the area of the Holmberg ellipse (corrected for inclination).  The observed \ha\ fluxes have to be corrected for contamination from NII, before they can be used for SFR estimation. For the low metallicity dwarfs in our sample, this correction is generally negligible. For three of our sample galaxies for which we have \ha\ observations, the NII flux is available separately. As expected the corrections are small ($\sim 0.01$~dex in \ssfr). We hence use the NII flux corrected data whenever possible (some literature \ha\ values are already corrected for NII flux contamination, as indicated in Table~\ref{tab:flux}) and ignore the correction otherwise.

\begin{table*}
\caption{Measured fluxes and literature sources}
\label{tab:flux}
\begin{tabular}{|lccccc|}
\hline
Galaxy&L$_{\rm FUV}$&L$_{{\rm H}\alpha}$&Reference&Reference&Reference\\
      &(ergs s$^{-1}$~Hz$^{-1}$)&(ergs s$^{-1}$)&\ha\ flux&NII flux&24\um\ flux\\
\hline
And IV$^{U~d}$        &${\rm 1.0\times10^{25}}$&                        &1    &~&~\\
DDO 226$^U$           &${\rm 2.1\times10^{25}}$&                        &2$^*$&~&~\\
DDO 6$^m$             &${\rm 8.3\times10^{24}}$&                        &3    &~&~\\
UGC 685$^{U}$         &${\rm 2.9\times10^{25}}$&${\rm 6.4\times10^{38}}$&~    &4&5\\
KK 14$^m$             &${\rm 7.1\times10^{24}}$&${\rm 3.7\times10^{37}}$&~    &~&~\\
KK 41$^m$             &${\rm 1.5\times10^{25}}$&${\rm 5.4\times10^{37}}$&~    &~&~\\
UGCA 92               &                        &                        &4$^*$&~&~\\
KKH 34                &                        &                        &1    &~&~\\
UGC 3755$^{U}$        &${\rm 1.1\times10^{26}}$&${\rm 1.4\times10^{39}}$&~    &~&~\\
DDO 43$^m$            &${\rm 7.2\times10^{25}}$&                        &1    &~&~\\
KK 65$^{U}$           &${\rm 2.6\times10^{25}}$&${\rm 7.3\times10^{38}}$&~    &~&~\\
UGC 4115$^{U~d}$      &${\rm 1.3\times10^{26}}$&                        &1    &~&~\\
KDG 52$^m$            &${\rm 5.7\times10^{24}}$&                        &~    &~&5\\
UGC 4459$^{U}$        &${\rm 3.5\times10^{25}}$&${\rm 1.0\times10^{39}}$&~    &4&5\\
UGC 5186$^U$          &${\rm 8.9\times10^{24}}$&${\rm 2.4\times10^{36}}$&~    &~&~\\
UGC 5209$^U$          &${\rm 1.1\times10^{25}}$&                        &1    &~&~\\
UGC 5456$^{U}$        &${\rm 1.2\times10^{26}}$&                        &1    &~&5\\
UGC 6145$^U$          &${\rm 1.3\times10^{25}}$&                        &~    &~&~\\
UGC 6456$^{U}$        &${\rm 7.9\times10^{25}}$&${\rm 2.0\times10^{39}}$&~    &4&6\\
UGC 6541$^m$          &${\rm 4.6\times10^{25}}$&                        &4$^*$&~&5\\
NGC 3741$^{U~d}$      &${\rm 2.9\times10^{25}}$&                        &4$^*$&~&5\\
DDO 99$^m$            &${\rm 2.6\times10^{25}}$&                        &1    &~&5\\
E321$-$014$^m$        &${\rm 7.0\times10^{24}}$&                        &3    &~&5\\
UGC 7242$^{U~d}$      &${\rm 3.7\times10^{25}}$&                        &~    &~&5\\
CGCG 269$-$049$^{U~d}$&${\rm 1.6\times10^{25}}$&${\rm 1.5\times10^{38}}$&~    &~&~\\
UGC 7298$^{U~d}$      &${\rm 7.4\times10^{24}}$&                        &~    &~&~\\
KK 144$^{U~d}$        &${\rm 1.1\times10^{25}}$&${\rm 5.4\times10^{37}}$&~    &~&~\\
DDO 125$^{U}$         &${\rm 2.6\times10^{25}}$&${\rm 2.6\times10^{38}}$&~    &~&5\\
UGC 7605$^{U}$        &${\rm 3.7\times10^{25}}$&${\rm 2.5\times10^{38}}$&~    &~&5\\
GR8$^{U}$             &${\rm 1.5\times10^{25}}$&                        &1    &~&5\\
UGC 8215$^U$          &${\rm 6.0\times10^{24}}$&                        &1    &~&~\\
DDO 167$^{U~d}$       &${\rm 1.6\times10^{25}}$&                        &1    &~&~\\
KK 200$^U$            &${\rm 4.2\times10^{24}}$&                        &3    &~&~\\
E444$-$78$^U$         &${\rm 1.0\times10^{25}}$&                        &~    &~&~\\
UGC 8638$^{U}$        &${\rm 3.4\times10^{25}}$&${\rm 2.5\times10^{38}}$&~    &~&5\\
DDO 181$^{U~d}$       &${\rm 1.9\times10^{25}}$&${\rm 2.5\times10^{38}}$&~    &~&5\\
DDO 183$^U$           &${\rm 1.6\times10^{25}}$&                        &1    &~&5\\
UGC 8833$^U$          &${\rm 9.1\times10^{24}}$&                        &1    &~&~\\
KK 230                &${\rm 6.5\times10^{23}}$&                        &~    &~&~\\
DDO 187$^U$           &${\rm 7.6\times10^{24}}$&                        &1    &~&~\\
KK 246$^U$            &${\rm 2.5\times10^{25}}$&                        &~    &~&~\\
UGCA 438$^m$          &${\rm 1.3\times10^{25}}$&                        &1    &~&~\\
KKH 98$^{U~d}$        &${\rm 2.4\times10^{24}}$&${\rm 3.0\times10^{37}}$&~    &~&~\\
\hline
\end{tabular}
\begin{flushleft}
$^U$: Galaxies with `trustworthy' SFR$_{FUV}$~values (see Section~\ref{ssec:sftrac})\\
$^d$: fluxes summed over ellipse with axes increased by 10\% compared to the values listed in Table~\ref{tab:samp}\\
$^m$: galaxies with morphologically disturbed HI (see Section~\ref{ssec:HImorph})\\
$^*$: corrected for NII flux.\\
References-- 1: \citet{ken08}; 2: \citet{meu06}; 3: \citet{bou09}; 4: \citet{mou06}; 5: \citet{dal09}; 6: \citet{eng08}.
\end{flushleft}
\end{table*}

The luminosities measured for each sample galaxy is listed in Table~\ref{tab:flux}, FUV in column (2) and \ha\ in column (3).  The sources for additional data taken from the literature are also given in Table~\ref{tab:flux}.
Columns (4), (5) and (6) provide the references from where values of \ha\ flux, NII flux and 24 \um\ flux were obtained.

The calibration we use for converting FUV luminosity to SFR is taken from \citet{ken12,hao11}:
\begin{equation}
\rm {log~\frac{SFR}{M_{\odot}~yr^{-1}}~=~log~\frac{\nu}{Hz} \frac{L_{\nu}}{ergs~s^{-1}~Hz^{-1}}~-~43.35}
\label{eqn:3_3}
\end{equation}
The calibration for converting \ha\ luminosity to SFR is also taken from \citet{ken12,hao11}:
\begin{equation}
\rm {log~\frac{SFR}{M_{\odot}~yr^{-1}}~=~log~\frac{L_{H\alpha}}{ergs~s^{-1}}~-~41.27}
\label{eqn:3_4}
\end{equation}

These updated calibrations use a Kroupa IMF with mass limits of 0.1 and 100 \ms, with a slope of $-$2.35 between 1 and 100 \ms\ and a slope of $-$1.3 between 0.1 and 1 \ms. The calibrations use the latest Starburt99 codes of \cite{lei99}, but are appropriate for solar metallicity. Our sample galaxies have estimated metalicities significantly lower than solar, and the correction for this is discussed below.

We account for the energy from star formation re-radiated at  infra-red wavelengths due to dust using {\it Spitzer} 24 \um\ data and the new `composite' calibrations. For star formation estimated from FUV we use the relation given by \citet{hao11}, viz.
\begin{equation}
\rm {L_{FUV,corr}~=~L_{FUV, obs}~+~3.89~L_{25~\mu m}}
\label{eqn:3_5}
\end{equation}
which is fed into equation~\ref{eqn:3_3} to obtain the SFR.

\begin{figure}
\begin{center}
\psfig{file=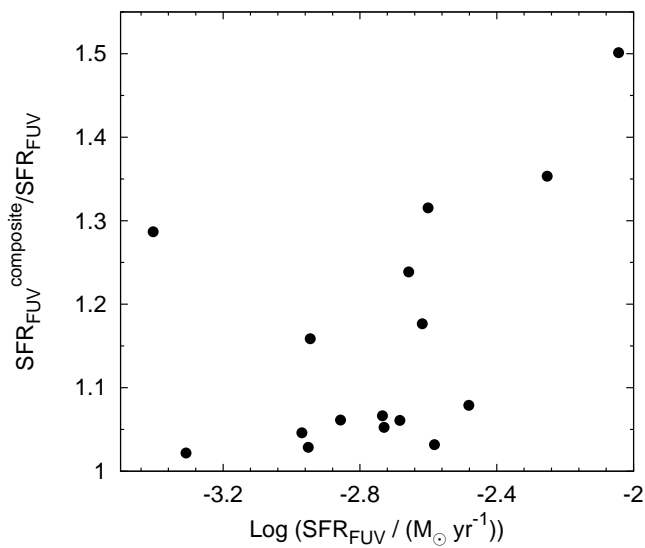,height=3.5truein,angle=270}
\end{center}
\caption{For the 16 galaxies with 24 \um\ data, ratio of the `composite' SFR calculated using eqn. 3 in eqn. 1 to the SFR calculated using eqn. 1 and the FUV flux only, plotted against the latter SFR in log.}
\label{fig:sfinc}
\end{figure}

For star formation estimated from the \ha\ flux we use the correction given by  \citet{ken09}, viz.
\begin{equation}
\rm {L_{H\alpha,corr}~=~L_{H\alpha,obs}~+~0.02~L_{25~\mu m}}
\label{eqn:3_6}
\end{equation}
which is fed into equation~\ref{eqn:3_4} to obtain the SFR.
Due to the low dust content of our sample dwarf galaxies, the correction from dust obscuration in small.
This is shown in Figure~\ref{fig:sfinc} where we plot the fractional change in SFR (accounted for the correction due to low metallicity discussed next) on accounting for FUV emission reprocessed by dust for sample galaxies with available 24 \um\ data.

\begin{figure*}
\begin{center}
\psfig{file=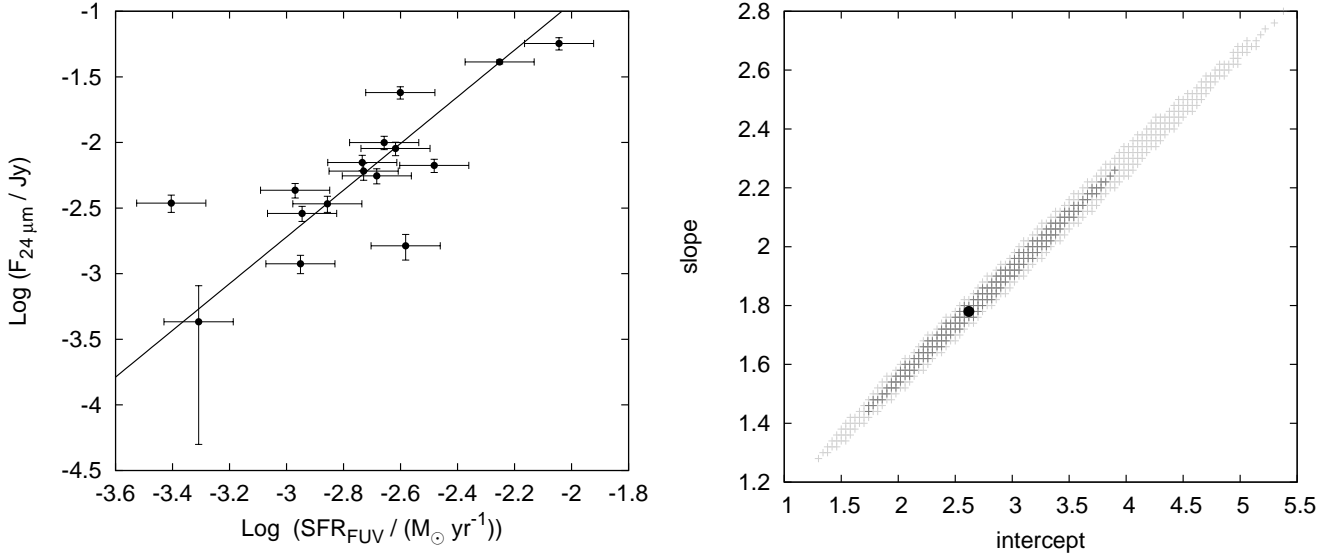,width=7truein}
\end{center}
\caption{Left panel shows the logarithm of measured 24 \um\ flux plotted against the logarithm of the total star formation rate within the Holmberg ellipse calculated using only FUV flux (and not the `composite' calibration) but after correcting for NII flux(wherever available) and the effect of sub-solar metallicity (see text for details), for sample galaxies with existing 24 \um\ measurements. The bold line shows the best fit to the points. The right panel shows the best fit (black point) value and 68\% (dark grey), 95\% (light grey) confidence intervals on the value of slope and intercept.}
\label{fig:irf}
\end{figure*}

\begin{figure*}
\begin{center}
\psfig{file=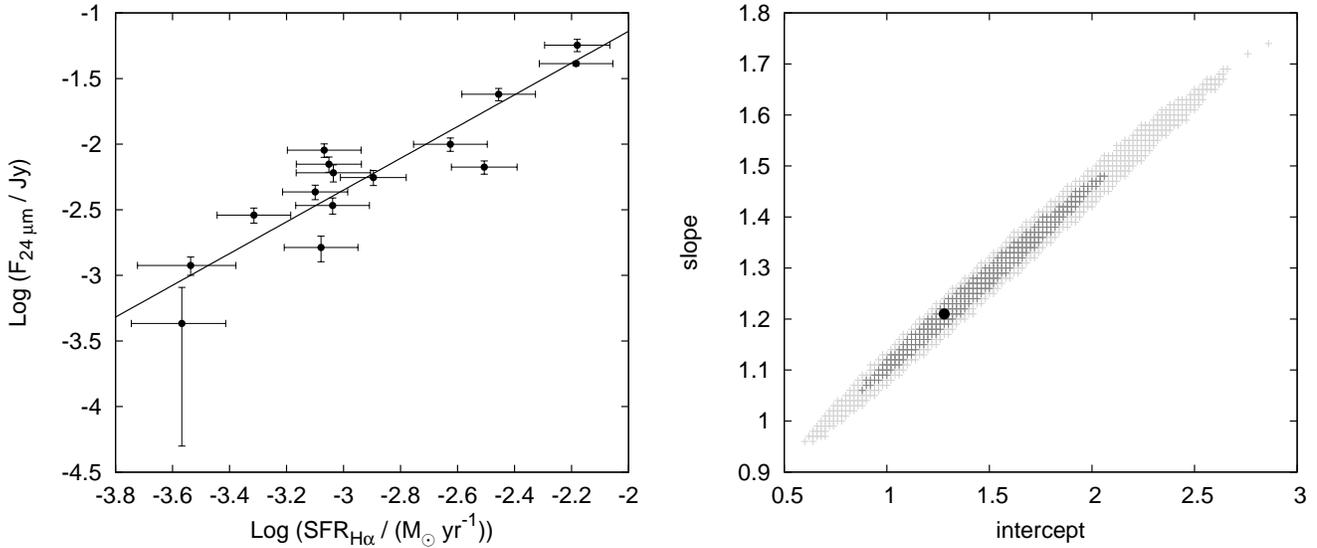,width=7truein}
\end{center}
\caption{Left panel shows the logarithm of measured 24 \um\ flux plotted against the logarithm of the total star formation rate within the Holmberg ellipse calculated using only \ha\ flux (and not the `composite' calibration) but after correcting for NII flux(wherever available) and the effect of sub-solar metallicity (see text for details), for sample galaxies with existing 24 \um\ measurements. The bold line shows the best fit to the points. The right panel shows the best fit (black point) value and 68\% (dark grey), 95\% (light grey) confidence intervals on the value of slope and intercept.}
\label{fig:irh}
\end{figure*}

{\it Spitzer} 24 \um\ fluxes are available only for some of the galaxies in our sample. In oder to estimate the dust correction for the other galaxies we use the available measurements to estimate how the 24 \um\ flux varies with the FUV flux. We show in Figures~\ref{fig:irf}~and \ref{fig:irh} the 24~\um\ flux against the SFR, determined without taking into account the IR emission but correcting for metallicity as described in the next paragraph, from FUV and \ha\ for galaxies with observed 24~\um\ fluxes. We fit power laws to both sets of data using the method described in Section~\ref{ssec:res}. The plots are shown in Figures~\ref{fig:irf}~and \ref{fig:irh} with the corresponding confidence intervals for the best fits, which are given by:
\begin{equation}
\begin{array}{l}
\displaystyle \rm{\log \frac{F_{24 \mu m}}{Jy}~=~1.78 \log \frac{SFR_{FUV}}{M_{\odot}~yr^{-1}}~+~2.62}\\\\
\displaystyle \rm{\log \frac{F_{24 \mu m}}{Jy}~=~1.21 \log \frac{SFR_{H\alpha}}{M_{\odot}~yr^{-1}}~+~1.28}
\end{array}
\label{eqn:irfh}
\end{equation}

For galaxies without 24 \um\ data, the expected 24 \um\ emission is estimated using the above mentioned best fits and the SFR of the galaxy. The estimated dust correction varies from 0.0007~dex to 0.4~dex with a median change of 0.05 dex for \ssfr\ measured using FUV , and varies from 0.006~dex to 0.23~dex with a median change of 0.09 dex for \ssfr\ measured using \ha. For most of the galaxies in our sample, this estimated correction is hence small.

Finally a correction is necessary in order to account for the low metallicities of our sample galaxies. The oxygen abundance of a handful of our sample galaxies has been measured by \citet{mar10}. We use the value of solar abundance from \citet{asp09} to arrive at the metallicity for these galaxies, and the resulting values are listedn in Table~\ref{tab:samp}. Additionally oxygen and Balmer band line fluxes for UGC 6456 is given in \citet{mou06}. We use this flux to derive abundance using the method of \citet{pil05} and hence  determine the gas phase metallicity, also listed in Table~\ref{tab:samp}. For the remaining galaxies without measured metallicities, we use the luminosity (M$_{\rm B}$) -- metallicity relation for dIs from \citet{ekt10} (the second relation in their Table 4) to estimate their metallicities. Recent estimates of emergent fluxes in sub-solar metallicity environments by \citet{rai10} calculated using evolutionary synthesis models using a Salpeter IMF and constant star formation for the last 10$^{\rm 8}$ years show that they increase by $\sim$ 11\%, 19\%, 27\%, 32\% and $\sim$ 18\%, 38\%, 67\%, 85\% for FUV and \ha\ ionizing fluxes respectively for metallicities of  0.4, 0.2, 0.05, 0.02 times solar. For each sample galaxy we do a linear interpolation between these values and arrive at the percentage increase and hence the correction factor for the emergent flux at the metalicity of that galaxy. Further, in order to account for variations in the IMF and star formation history, we increase the estimated error on these corrected \ssfr. Details on this are given in the following section. We use the metallicity corrected \ssfr\ for all further calculations.

\subsection*{Error estimates}
\label{ssec:ee}

We estimate the error on the measurement of \shi\ to be 10\%, which accounts for both the noise in the images as well as for errors in the flux calibration. For the \ssfr\ the errors are computed as the quadrature sum of whichever of the following errors are applicable for the galaxy under consideration: flux measurement errors (FUV,\ha,NII,24 \um), 10\% flux calibration error for {\it GALEX} FUV data, 15\% flux calibration error for BTA \ha\ data. Following \citet{ler12,ler13} an additional 50\% error is added in quadrature in order to account for the errors caused by variations in the IMF and star formation history. Whenever SFR was calculated without considering the contribution from 24 \um\ flux, the error on the calibration was taken to be 30\%. For galaxies without measured 24 \um\ fluxes, the error on the estimated 24 \um\ fluxes (i.e. those determined using the fits in Eqn.~\ref{eqn:irfh}) were computed using the 68\% confidence intervals shown in Figures~\ref{fig:irf}~and \ref{fig:irh}. These errors are also added in quadrature to arrive at the total estimated error.

\section{Results and Discussion}

\subsection{Comparing the star formation tracers}
\label{ssec:sftrac}

The tracers that we use here, viz. the \ha\ and FUV fluxes, are sensitive to different parts of the stellar IMF. \ha\ emission primarily traces massive ($\gtrsim$ 16 \ms) star formation while FUV emission arises from the photospheres of intermediate mass ($\gtrsim$ 6 \ms), longer lived stars. The FUV emission is also more affected by extinction by dust than \ha\ emission. These and other phenomena such as the escape of ionizing photons and possible variations in the IMF between galaxies \citep{wei05,meu09} can produce differences in the SFRs as estimated using the these two different tracers. Such differences are expected to be most pronounced in dwarf irregular galaxies with low overall SFRs. For example, a recent  burst of star formation (leading to significant \ha\ flux) superposed on a quiescent SFR (which is traced by FUV) can make the SFR estimated using \ha\ higher than that traced using FUV. Such  mismatches have been observed for several dwarf galaxies \citep[see e.g.][]{lee09,hun10}, even for some of the galaxies from the present sample \citep{roy11}. A recent work comparing the two SFRs for Local Volume dwarf galaxies which includes all our present sample galaxies observe a similar mismatch at the faint end \citep{kar13a}. Observations suggest that these difference are unlikely to be due to the escape of  ionizing photons \citep{rel12} or variation in the IMF \citep{roy11,wei12,her13}. Here we look at the SFRs obtained for our sample galaxies using \ha\ and FUV and see whether the measurements can be reconciled in terms dust attenuation, bursty star formation and stochastic sampling of the IMF.

\begin{figure}
\begin{center}
\psfig{file=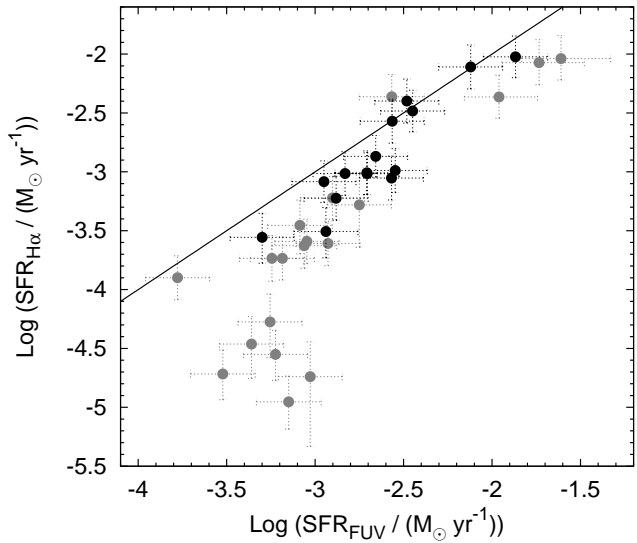,height=3.5truein,angle=270}
\end{center}
\caption{Logarithm of total SFRs within the Holmberg ellipse deduced using \ha\ emission and FUV flux plotted against one another. Galaxies with existing 24 \um\ measurements are represented with black circles while the ones without without with grey circles. The bold line is the 1:1 line.}
\label{fig:tsf}
\end{figure}

Figure~\ref{fig:tsf} shows the SFRs calculated using the two different tracers for our sample galaxies plotted against each other. 
The most noteworthy feature of Fig.~\ref{fig:tsf} is that at low star formation rates (i.e. for $\rm{log(SFR/(M_{\odot}~yr^{-1}))}~\lesssim~\rm{-3.0}$), the SFR estimated from the \ha\ flux is systematically smaller than that estimated from the FUV flux. \citet{daS14} study the stochastic effects at low intrinsic SFRs and how they affect SFR estimates derived using various tracers. They find that as one measures increasingly low SFRs using a particular tracer through a standard calibration function, the posterior probability distribution function (PDF) of the true underlying SFR not only becomes wider (due to the stochastic sampling of the high mass end of the IMF) but also its peak is offset to a higher SFR value than the one measured (due to the increasingly bursty nature of star formation at low SFRs).
This is most pronounced when SFR is traced by \ha\ emission.
Using the output of their code available online, we study the behaviour of the posterior SFR PDF when varying the SFR as measured using \ha\ or FUV in steps of 0.25 dex. We consider their results for a flat prior distribution of the SFR and with the assumption that all star formation occurs in either clusters or associations, consistent with the increasingly bursty star formation histories with decreasing stellar mass in galaxies \citep{kau14}. When using \ha\ as tracer the peak of the posterior PDF is already offset from the measured SFR by more than 0.25 dex for the highest SFRs measured in our sample galaxies, and the offset increases to $\sim$0.5 dex for $\rm{log(SFR/(M_{\odot}~yr^{-1}))}~\lesssim~$ -2.4, and to $\sim$0.75 dex for $\rm{log(SFR/(M_{\odot}~yr^{-1}))}~\lesssim~-$ 3.4. The scatter in the posterior SFR PDF is also substantial and increases to more than 1 dex for $\rm{log(SFR/(M_{\odot}~yr^{-1}))}~\lesssim~-$ 2.5. Considering the above facts, we choose $\rm{log(SFR/(M_{\odot}~yr^{-1}))}~=~-$ 2.5 as the limiting trustworthy SFR measured using \ha\ emission. When using FUV as a tracer of star formation though, the situation is markedly better.
The peak of the posterior SFR PDF shifts by $\sim$0.25 dex as compared to the measured SFR only for $\rm{log(SFR/(M_{\odot}~yr^{-1}))}~\lesssim~-$ 2.8, and by $\sim$0.5 dex for $\rm{log(SFR/(M_{\odot}~yr^{-1}))}~\lesssim~-$ 4.9.
The scatter in the posterior SFR PDF remains low and approaches 1 dex only for $\rm{log(SFR/(M_{\odot}~yr^{-1}))}~\lesssim~-$ 4.
Considering the above mentioned values, it becomes obvious that \ssfr s obtained for our sample galaxies using FUV as tracer are more trustworthy.
We use $\rm{log(SFR/(M_{\odot}~yr^{-1}))}~=~-$ 4 as the limiting trustworthy SFR measured using FUV emission.

\subsection{Morphology of the HI distributions}
\label{ssec:HImorph}

Having established the regions where the star formation indicators appear to be reliable, we now move on to comparing  star formation rate density to the gas surface density. For spiral galaxies, (as detailed in the introduction), these two surface densities are found to to be correlated.  The situation in very small dwarf galaxies could however be more complex. The shallow potential wells of dwarf galaxies makes it easy for feedback from star formation to have a significant impact on the distribution and kinematics of their gaseous component \citep[e.g. see][]{mac99,gov10}, but see also \citet{sil01}. In the context of the current study, the principal effect of this feedback is that it could destroy any previously extant correlations between the gas density and the star formation rate. We hence carefully examine all of the HI distributions in our galaxy sample to identify those for which there is a clear signature of the HI gas being disturbed. For 9/43 galaxies (viz. DDO~43, KDG~52, DDO~6, KK~14, KK~41, UGC~6541, DDO~99, E321-014 and UGCA 438) we find that the HI distribution is disturbed.  We exclude these 9 galaxies from the  analysis below where we compare the gas and star formation surface densities. The ``disturbances'' vary from the galaxy having an HI ``hole'' at the center of the disk (e.g. DDO~43 and KDG~52), to galaxies where the HI is misaligned or even non-overlapping with the star forming disk (Figure~\ref{fig:ov}). The fact that about one-fourth of our sample galaxies show {\it morphological} signatures of disturbed HI distributions is an indication that star formation feedback could be important in dwarf galaxies.

\subsection{The relation between \sgas\ and \ssfr}
\label{ssec:res}

\begin{figure}
\begin{center}
\psfig{file=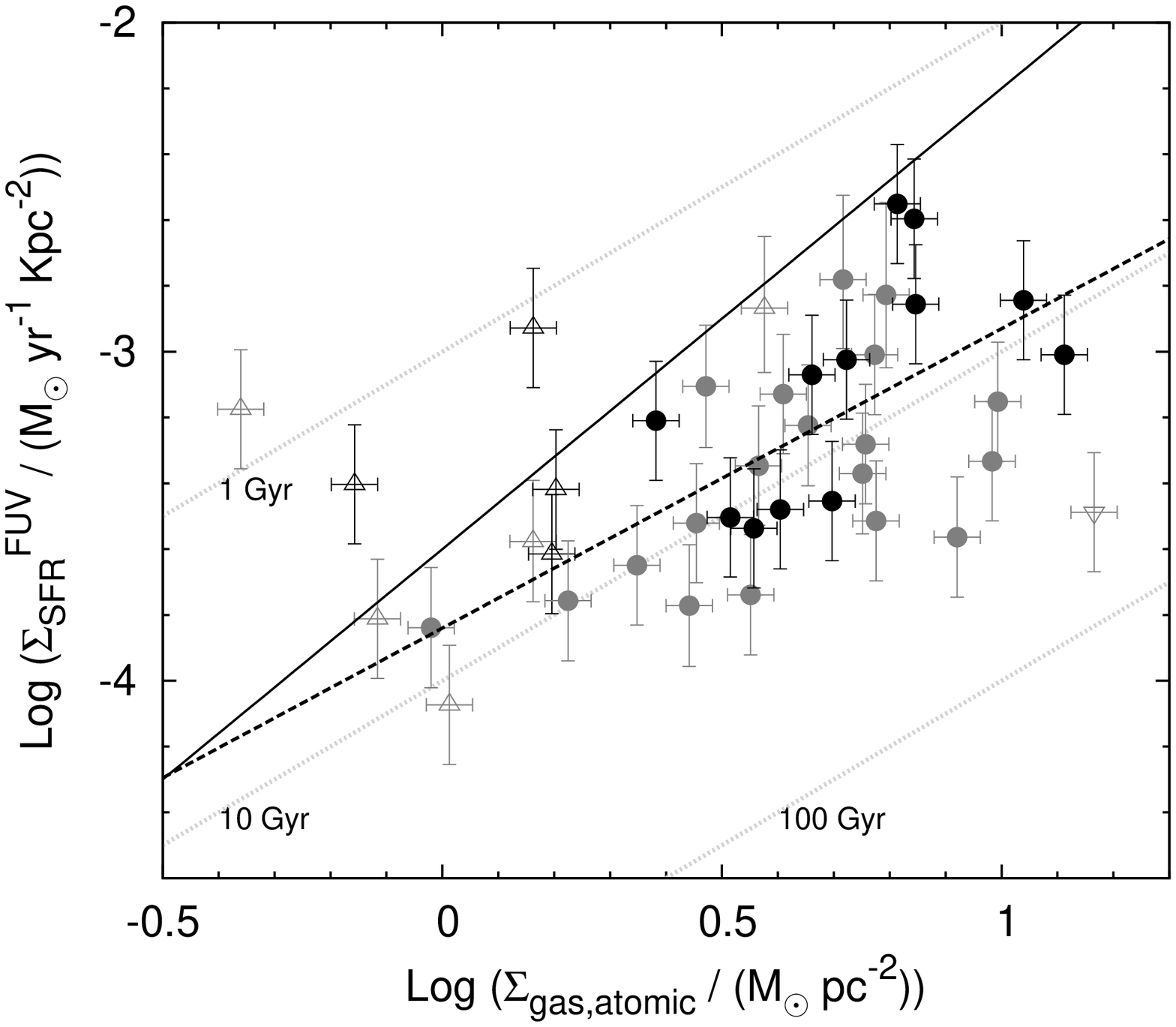,height=3.5truein,angle=270}
\end{center}
\caption{Logarithm of disk-averaged surface densities of SFR (deduced using FUV flux) and atomic gas plotted against each other. Black symbols are for galaxies with existing 24 \um\ measurements and grey symbols are for those without. Open upward pointing triangles represent galaxies with disturbed HI morphology whereas open downward pointing triangle represents the only sample galaxy with total SFR below $-$4 in log. The remaining galaxies are marked with filled circles. The best fit Schmidt law to the remaining galaxies marked with filled circles is shown as the dashed line, whereas the `canonical' Kennicutt-Schmidt law is shown as the bold line. Faint grey dotted lines mark different gas consumption timescales.}
\label{fig:kslawf}
\end{figure}

\begin{figure}
\begin{center}
\psfig{file=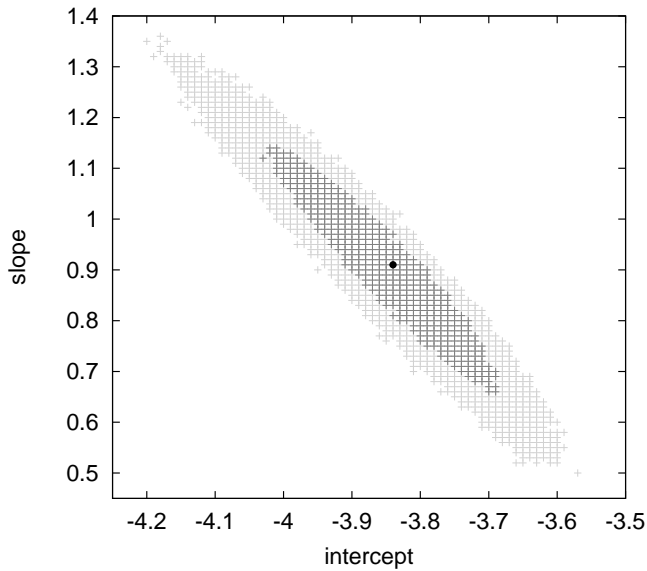,height=3.5truein,angle=270}
\end{center}
\caption{Corresponding to the fit shown in Figure~\ref{fig:kslawf}, the best fit (black point) value and 68\% (dark grey), 95\% (light grey) confidence intervals on the value of slope and intercept are shown.}
\label{fig:ci}
\end{figure}

\begin{figure}
\begin{center}
\psfig{file=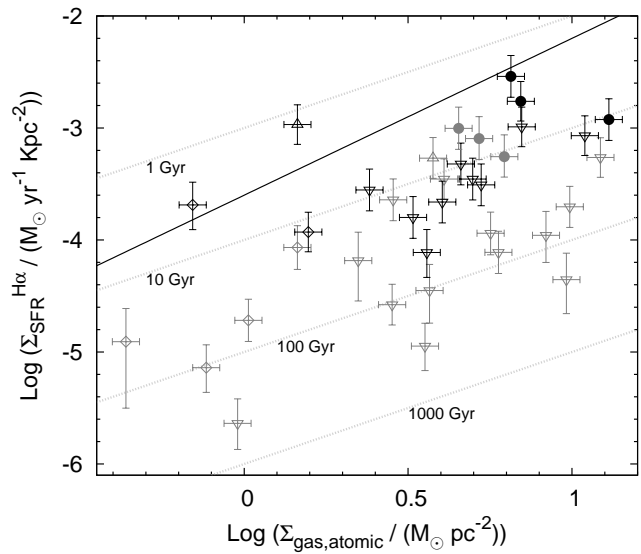,height=3.5truein,angle=270}
\end{center}
\caption{Logarithm of disk-averaged surface densities of SFR (deduced using \ha\ flux) and atomic gas plotted against each other. Black symbols are for galaxies with existing 24 \um\ measurements and grey symbols are for those without. Open upward pointing triangles represent galaxies with disturbed HI morphology, open downward pointing triangles represent galaxies with total SFR lower than the limit below which there would be too few \ha\ producing stars whereas open diamonds represent galaxies which have both disturbed HI morphology and total SFR below the `trustworthy' threshold. (see text for details). The remaining galaxies are marked with filled circles. Points with dashed errorbars represent galaxies with literature \ha\ data. The `canonical' Kennicutt-Schmidt law is shown as the bold line. Faint grey dotted lines mark different gas consumption timescales.}
\label{fig:kslawh}
\end{figure}

Figures~\ref{fig:kslawf} and \ref{fig:kslawh} show scatter plots of \ssfr\ and \sgas\ for the galaxies in our sample. Galaxies for which the SFR is low enough for stochastic effects to be important (see discussion in Section~\ref{ssec:sftrac}) are marked with downward pointing triangles while galaxies with disturbed HI morphology are marked by upward pointing triangles. 
Two things stand out for these galaxies in line with our expectations, viz. (i) for galaxies with disturbed HI morphology the inferred gas consumption timescales are lower than the average for the sample indicating deficiency (plausibly loss) of HI within the star-forming region. And (ii) for galaxies below the SFR limit for the particular tracer the inferred gas consumption timescales are higher than the average for the sample indicating that the SFR has been underestimated.

In the case of \sfh\ excluding galaxies with low SFR and/or disturbed HI morphology leaves us with too few (six) galaxies to try and fit a Kennicutt-Schmidt type power law. Nevertheless, it is interesting to note that the gas consumption timescales for these galaxies clusters around $\sim$10 Gyr. This is about an order of magnitude shorter than that estimated for the molecule poor outer parts of disk galaxies \citep{big10}. For the galaxies with SFR rate estimated from the FUV emission we are left with 31 galaxies after removing galaxies with low SFR and/or disturbed HI. We refer to this sub-sample as the `trustworthy' sub-sample. For these galaxies we determine the best fit linear relation through a Monte-Carlo method which is similar to a direct bivariate linear regression but at the same time provides a way to estimate the error on the fitted quantities notwithstanding the asymmetric error bars in our data. 

For each `trustworthy' galaxy, the surface density (HI or SFR) is assumed to have a distribution which is a normalized combination of two Gaussians having mean equal to the actual measurement. Values greater than the actual measurement are drawn from a Gaussian having standard deviation equal to the positive error bar (in real space), whereas values less than the actual measurement are drawn from a Gaussian having standard deviation equal to the negative error bar. 10$^{\rm 6}$ sets of data are simulated, where each set contains one value for each `trustworthy' galaxy drawn randomly from its \shi\ and \ssfr\ distributions defined in the above manner.
Straight lines are fitted to each of the 10$^{\rm 6}$ sets of data through bivariate linear regression, with the point for any galaxy weighted by the quadrature mean of the higher and lower error bars.
The peak of the 10$^{\rm 6}$ straight line fits (the best fit) is shown as the dashed line in Figure~\ref{fig:kslawf}, and also as the black point in Figure~\ref{fig:ci}.
Figure~\ref{fig:ci} also shows the 68\% and 95\% confidence intervals for the values of slope and intercept thus determined as the dark and light grey shaded areas respectively.
Using the extent of the 68\% confidence interval, the Kennicutt-Schmidt law using only atomic gas for the faint dwarf irregular galaxies in our sample is given as:
\begin{equation}
\rm{\log \Sigma_{SFR} = 0.91^{+0.23}_{-0.25} \log \Sigma_{gas,atomic} - 3.84^{+0.15}_{-0.19}}
\label{eqn:3_7}
\end{equation}

\noindent
In order to provide a straightforward comparison with \cite{ken98} we also determine the regression relation using the same procedure as followed in that paper. This recovers the same value for the mean slope (0.91) and almost the same value for the mean intercept ($-$3.87) as above.
 
\begin{figure*}
\begin{center}
\psfig{file=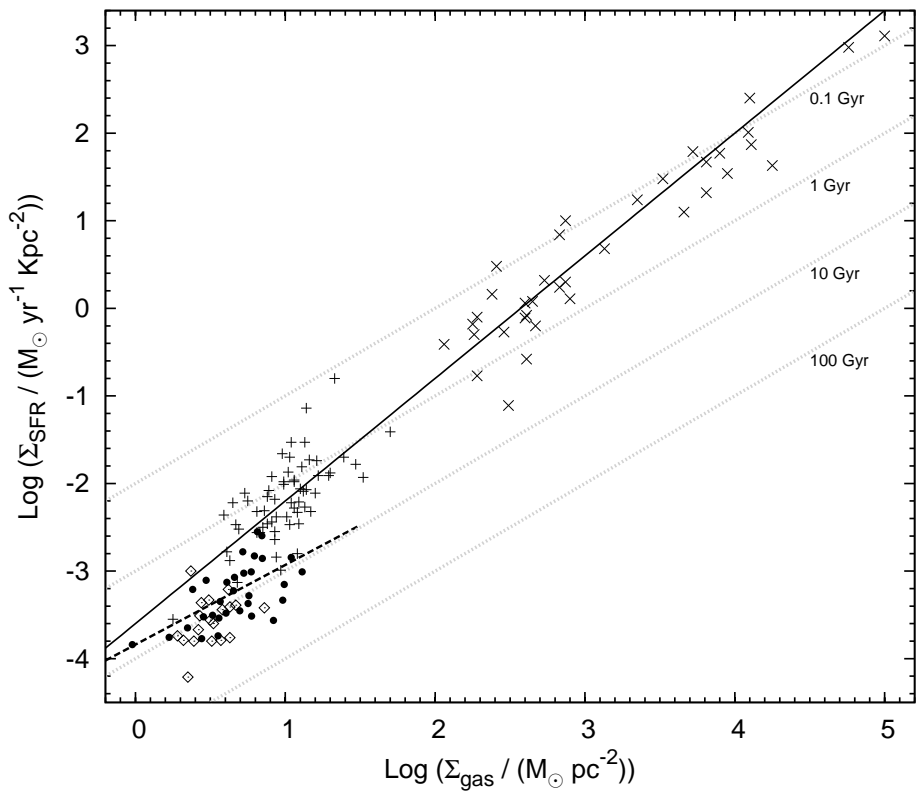,height=5truein,angle=270}
\end{center}
\caption{The `canonical' disk-averaged Kennicutt-Schmidt law compared to our results. Crosses and pluses represent circumnuclear starbursts and spirals from \citet{ken98} respectively. Open diamonds represent LSB galaxies from \citet{wyd09}. Our sample galaxies (SFR estimated using FUV emission) are represented by filled circles, without errorbars for clarity. The `canonical' Kennicutt-Schmidt law is shown as the bold line. The best fit Schmidt law to our sample galaxies is shown as the dashed line. Faint grey dotted lines mark different gas consumption timescales.}
\label{fig:kslawa}
\end{figure*}

In Figure~\ref{fig:kslawa} we compare the data for our galaxies (i.e. the `trustworthy' sub-sample) with that for other samples. The data on circumnuclear starbursts and spirals are from \citet{ken98}, and the are the ones based on which the `canonical' Kennicutt-Schmidt law was defined. Values for low surface brightness (LSB) galaxies are from \citet{wyd09}, who also estimate the gas surface density only using atomic gas. It is interesting to note that the LSB galaxies overlap with our sample galaxies albeit with a larger scatter. The fit to our sample galaxies also appears to be a good fit for LSB galaxies.
From the figure (see also Sec.~\ref{sec:totrel}) one can see that  our the gas consumption timescales for are sample galaxies is $\sim 10$~Gyr. This is significantly smaller than the estimated timescales  \citep[$\sim 100$~Gyr,][]{big10} for the outskirts of spiral galaxies. It is also significantly larger than  the estimated gas consumption timescales ($\sim 2$~Gyr ) in the inner parts of spiral galaxies \citep[see e.g.][]{ler13}.  

\begin{figure}
\begin{center}
\psfig{file=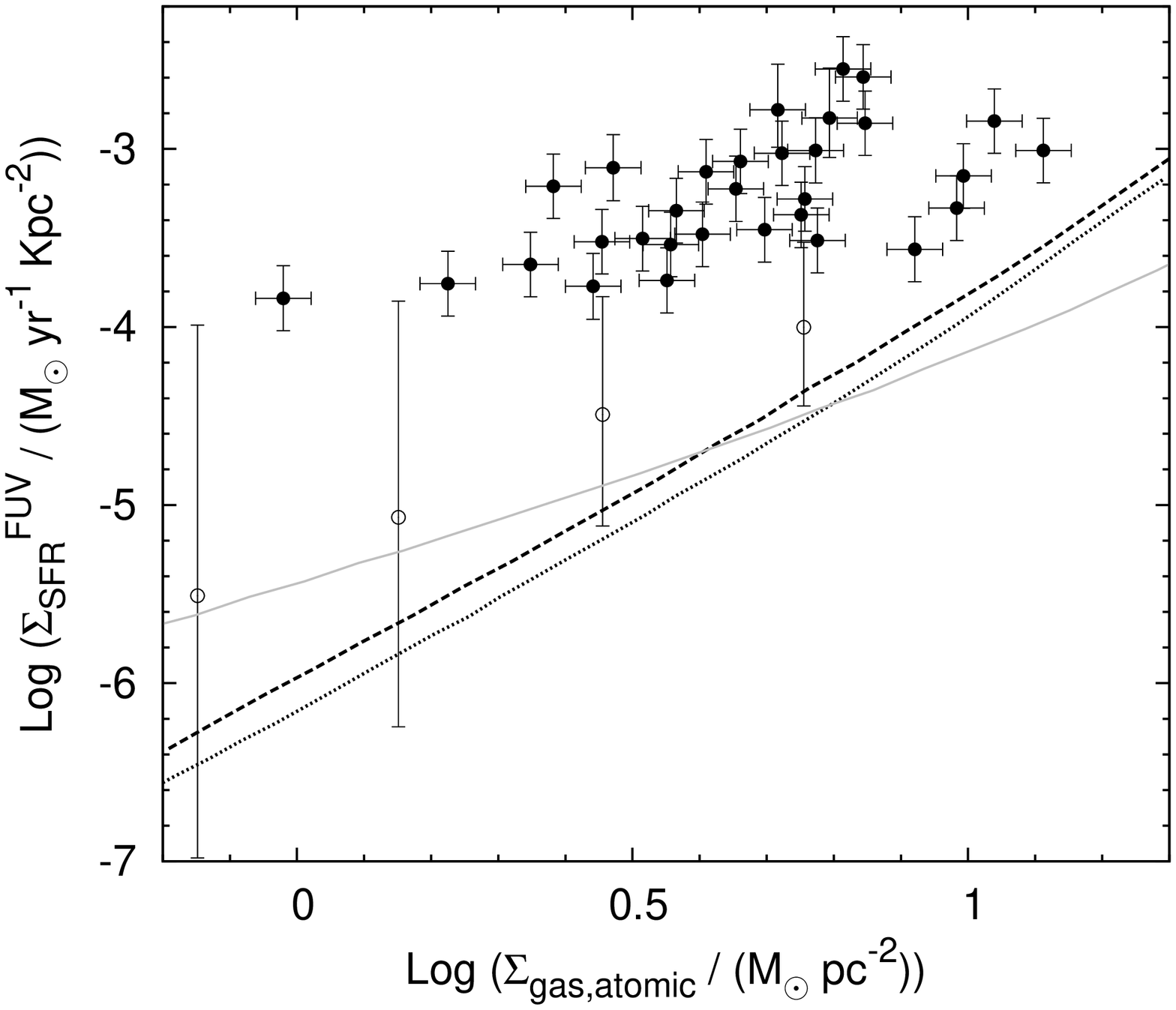,height=3.5truein,angle=270}
\end{center}
\caption{Data for the `trustworthy' galaxies in our sample is shown as the filled circles.The hollow circles with error bars show the median and scatter of $\sim$ kpc size regions in outskirts of spiral galaxies from \citet{big10}. The dashed and dotted lines show the expected variation of \ssfr\ with \sgas\ from \citet{kru13} with the model parameters being $f_c = 5,\rho_{sd} = 0.01,f_w = 0.5$ and $f_c = 5,\rho_{sd} = 0.03,f_w = 0.07$ respectively. The grey line shows the expected variation of \ssfr\ with \sgas\ according to the modification of the \citet{ost10} model in \citet{bol11} with the model parameter $\rho_{sd} = 0.01$. All the three models plotted here are for 0.1 times the solar metallicity (see text for details).}
\label{fig:k13}
\end{figure}

Ideally, one would like to look at the relation between the molecular gas and star formation, as opposed to the atomic gas and star formation. Unfortunately, it is not possible to do such a study for the faint dwarfs, since there is essentially no data on their molecular gas content.  Determination of the molecular hydrogen fraction remains difficult for low metallicity galaxies like the ones in our present sample due to the very high (expected) CO-to-H$_{\rm 2}$ conversion factors \citep{bol13}. CO measurements for galaxies with metallicities comparable to the upper range of metallicities for our sample galaxies are only now being done \citep{elm13}, though the Atacama Large Millimeter Array (ALMA) might change that soon. In general however, one would expect the molecular gas content of low mass dwarf irregular galaxies to be low.  Models for star formation in such environments have been presented by \citet{kru13}. \citet{ost10} also present models for star formation in regions of low \ssfr.  The model by \citet{ost10} assumes that the atomic gas has achived two phase thermal equilibrium while in \citet{kru13} model, the  formation of the CNM phase of atomic gas as well as the atomic to molecular transition and the star formation rate are set by the requirements of hydrostatic balance and not by two-phase equilibrium. \citet{bol11} have proposed a modification to the \citet{ost10}  model which brings it into  agreement with the data for the low metalicity conditions of the SMC. We compare these models with our data in Fig.~\ref{fig:k13}. In the case of \citet{kru13} model, we use the model with metallicity $\sim$ 0.1 times the solar metallicity, (which corresponds well to the estimated metallicity of our sample galaxies) with clumping factor $f_c = 5$ since for our sample galaxies we average flux over $\sim$ kpc size star-forming disks. The dashed and dotted lines are for $\rho_{sd} = 0.01, f_w = 0.5$, and $\rho_{sd} = 0.03, f_w = 0.07$. $\rho_{sd}$ is the volume density of stars and dark matter, (in units of $\rm{M_{\odot}~pc^{-3}}$) and the range used above corresponds well to the expected range for our sample galaxies. $f_w$ is a measure of the distribution of atomic gas in the different phases, and \citet{kru13} argues that the values used above bracket what one would expected in astrophysical situations. The solid grey line is the \citet{bol11} modification of the \citet{ost10} model with metallicity $\sim$ 0.1 times the solar metallicity and $\rho_{sd} = 0.01$. As can be seen, both models significantly under-predict the observed star formation rate for our sample galaxies. It is worth noting however that the points from \citet{big10} lie within the region where stochastic effects would lead to significant uncertainty in the estimated star formation rate (see the discussion in Section.~\ref{ssec:sftrac}).

As mentioned above, the molecular gas content of our galaxies is currently unknown. If we assume that the star formation rate is an indicator of the molecular gas content, then we can turn the question around, and try and estimate the molecular gas fraction in our sample galaxies from the observed \ssfr. In the outskirts of disk galaxies at least, this approach would be reasonable -- \citet{sch11} show that even in the atomic gas dominated outskirts of the disk galaxies, the \ssfr\ -- \shtwo\ relation is similar to that in the molecule rich regions. Two recent studies, viz. \citet{ler13} and \citet{mom13} propose significantly different estimates of the relation between \ssfr and \shtwo. To explore the range of possibilities we use both of these estimates. From the relationship given by \citet{ler13} we get a typical molecular fraction f$_{H_2} \sim 0.05$ for our sample galaxies, while the somewhat steeper relation given in \citet{mom13} gives f$_{H_2} \sim 0.4$. A molecular fraction of $\sim 0.4$ appears somewhat large for our sample galaxies considering the non-detection of CO emission even in the most luminous of our sample galaxies \citep[see][]{tay98,ler05,buy06,sch12}. However even the lower value estimated from the \cite{ler13} relation is $\sim 2$ times larger than that estimated for WLM, which has $Z \sim 0.13$ \cite{elm13}. 

\subsection{Comparing total HI available to the total SFR}
\label{sec:totrel}

\begin{figure}
\begin{center} 
\psfig{file=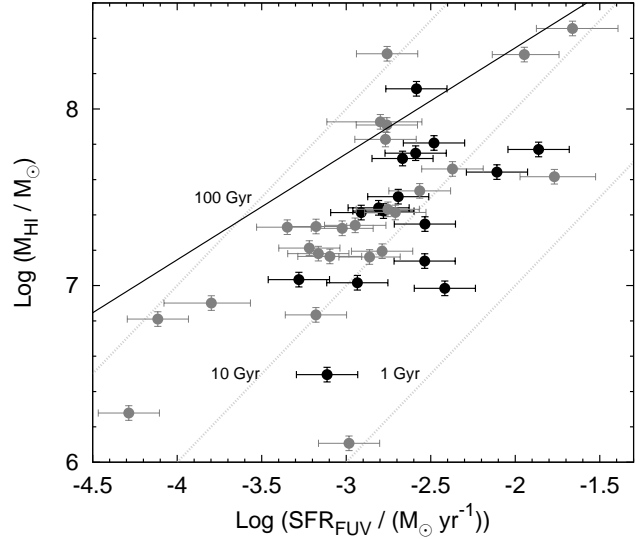,height=3.5truein,angle=270}
\end{center}
\caption{Total mass of HI plotted against to the total star formation rate of galaxies with (black points) and without (grey points) infrared data. The bold line is the best fit relation from \citet{doy06} for HI flux selected galaxies more massive than our sample (see text for details). The dotted lines indicate the loci for different gas consumption timescales.}
\label{fig:tot}
\end{figure}

We restricted our study of SFR and gas surface densities to within the `star-forming disk' of galaxies without any obvious signs of their HI disks being morphologically disturbed, for reasons described previously. But using this strategy meant we were discounting the (sometimes considerable) amount of HI present outside the `star-forming disk' (see Figure~\ref{fig:ov}) and its potential to be a fuel for star formation. We were also missing out on the relation between gas and star formation in the nine galaxies with morphologically disturbed HI. To address these issues, we do a comparison of the total star formation rate and the total HI mass of our sample galaxies in Figure~\ref{fig:tot}. The total HI masses for all the galaxies except UGCA 438 are taken from \citet{beg08} which were calculated using the coarsest resolution (5 K$\lambda$) HI data cubes. For UGCA 438 we calculated the total HI mass from the 400 pc resolution HI map used in this study.
We use all the 41 galaxies in our sample with FUV data for we calculate the total SFRs using FUV, the more trustworthy tracer of star formation for low SFRs. 
\citet{doy06} did a similar comparison for HI selected nearby galaxies from the HIPASS survey withe HI masses somewhat larger than that of our sample galaxies (i.e. ranging from a few times ${\rm 10^8~M_{\odot}}$ to a few times ${\rm 10^{10}~M_{\odot}}$.)  They found a power law relation between total SFR and HI mass with a slope of $\sim$0.6, this is shown in Figure~\ref{fig:tot} as a solid line. Our sample galaxies tend to lie below this line. This means that while dwarf galaxies do convert their gas to stars somewhat less efficiently than spirals, the process is not as inefficient as the extrapolation of the \citet{doy06} relationship would suggest. From the total HI mass and total SFR, the mean (median) gas consumption timescales for our sample galaxies is $\sim 18 (24)$~Gyr. While this is about a factor of $\sim 2$ larger than the values estimated using the gas within the star forming disk, it is still significantly smaller than the estimated timescales \citep[$\sim 100$~Gyr,][]{big10} for the outer parts of disk galaxies.

\section{Summary}
\label{ssec:dis}

We compare the global average star formation rate density \ssfr, as measured using different tracers with the average atomic gas surface density $\Sigma_{gas,atomic}$ for a sample of dwarf galaxies drawn from the FIGGS survey.  We use the \ha\, FUV and 24 \um\ fluxes to estimate \ssfr. The differences between the \ssfr\ computed with and without corrections for dust are small for most of the galaxies in our sample, consistent with their low metallicity. Excluding galaxies with  \ssfr\ too low to be reliably measured, as well as galaxies with disturbed HI distributions, we find a nearly linear relation between \ssfr\ and \sgas\, with a gas consumption time scale of $\sim 10$~Gyr. The typical gas consumption timescales of the star forming disks of dwarf galaxies is hence intermediate between that in the inner molecule rich and the outer molecule poor regions of spiral galaxies.

\section*{Acknowledgments}
Some of the data presented in this paper were obtained from the Multimission Archive at the Space Telescope Science Institute (MAST). STScI is operated by the Association of Universities for Research in Astronomy, Inc., under NASA contract NAS5-26555. Support for MAST for non-HST data is provided by the NASA Office of Space Science via grant NAG5-7584 and by other grants and contracts.
We thank the staff of the GMRT who have made the observations used in this paper possible.
GMRT is run by the National Centre for Radio Astrophysics of the Tata Institute of Fundamental Research.
SSK and IDK acknowledge that this work was supported by the Russian Foundation for Basic Research through grant 13-02-92690.
We thank Ayesha Begum for providing the visibilities which were used to derive the HI maps used in this work.
SR thanks Maryam Arabsalmani for helping with the Monte-Carlo simulations for fitting data with asymmetric errorbars.
SR thanks Guinevere Kauffmann for helpful discussions regarding this work.

\appendix
\section{Visual comparison of HI and star formation}
\label{sec:ap}

\begin{figure*}
\begin{center}
\begin{tabular}{cccc}
UGC 685&UGC 3755&KK 65&UGC 4459\\
{\mbox{\includegraphics[height=4cm,angle=270]{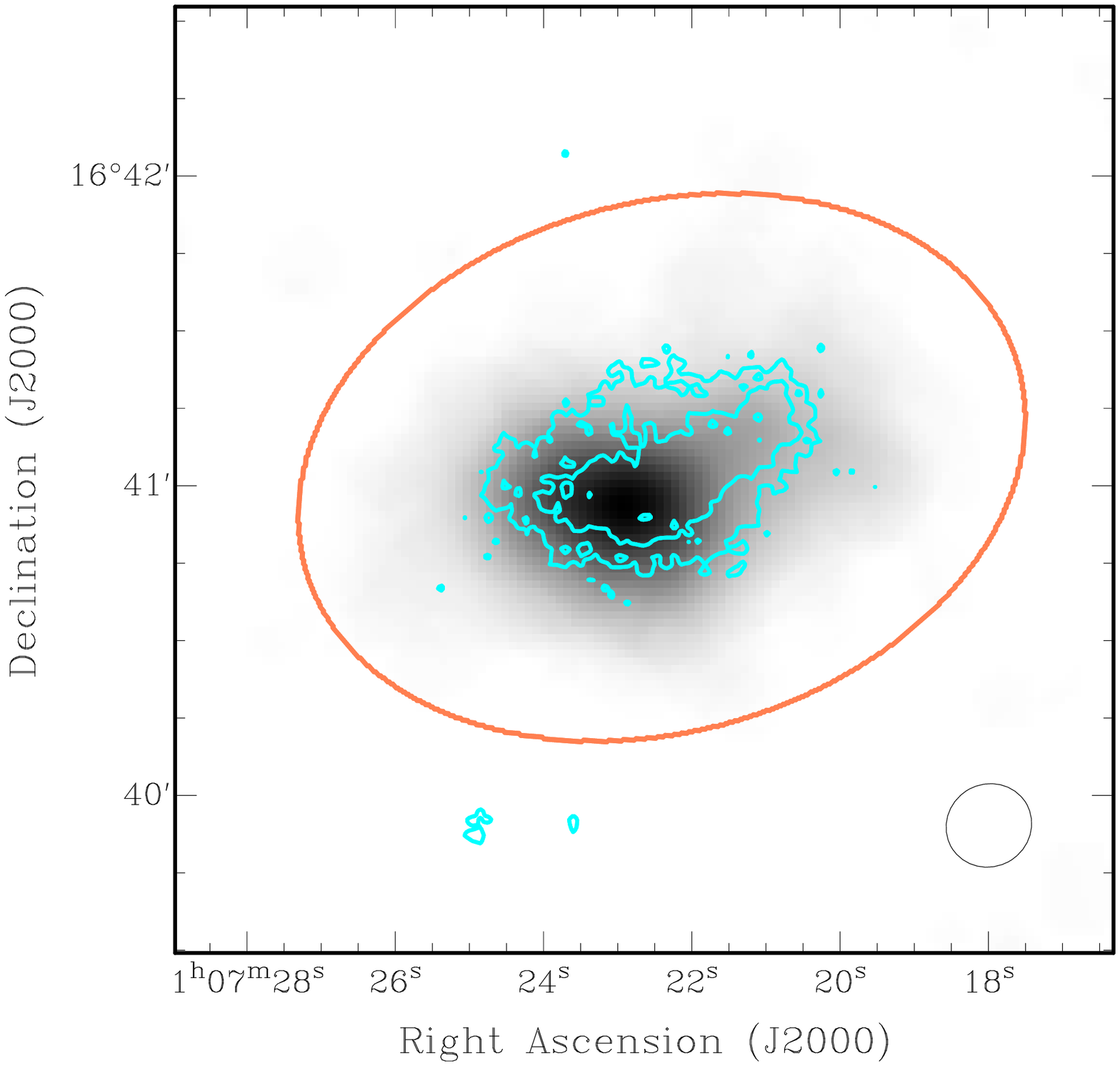}}}&
{\mbox{\includegraphics[height=4cm,angle=270]{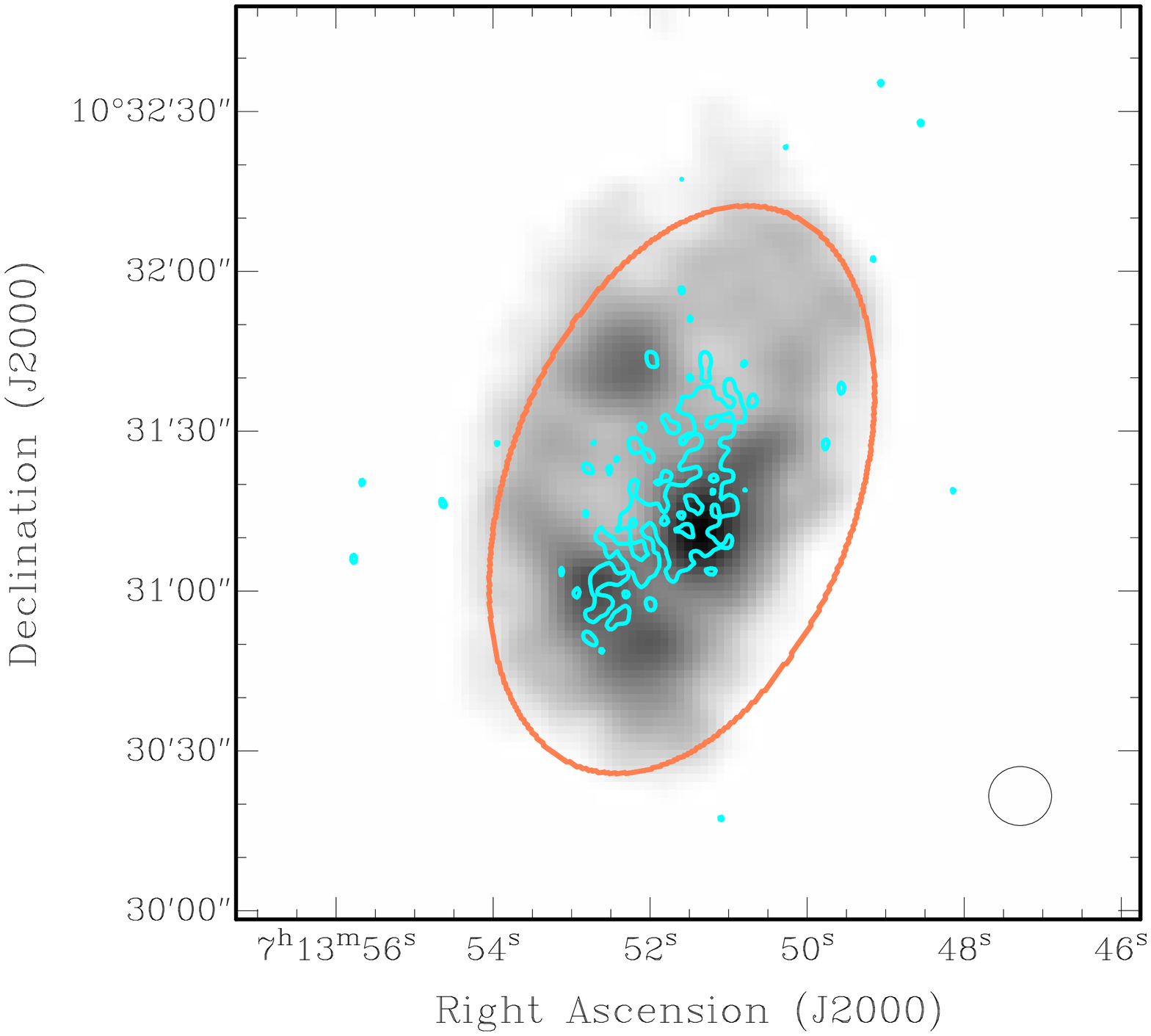}}}&
{\mbox{\includegraphics[height=4cm,angle=270]{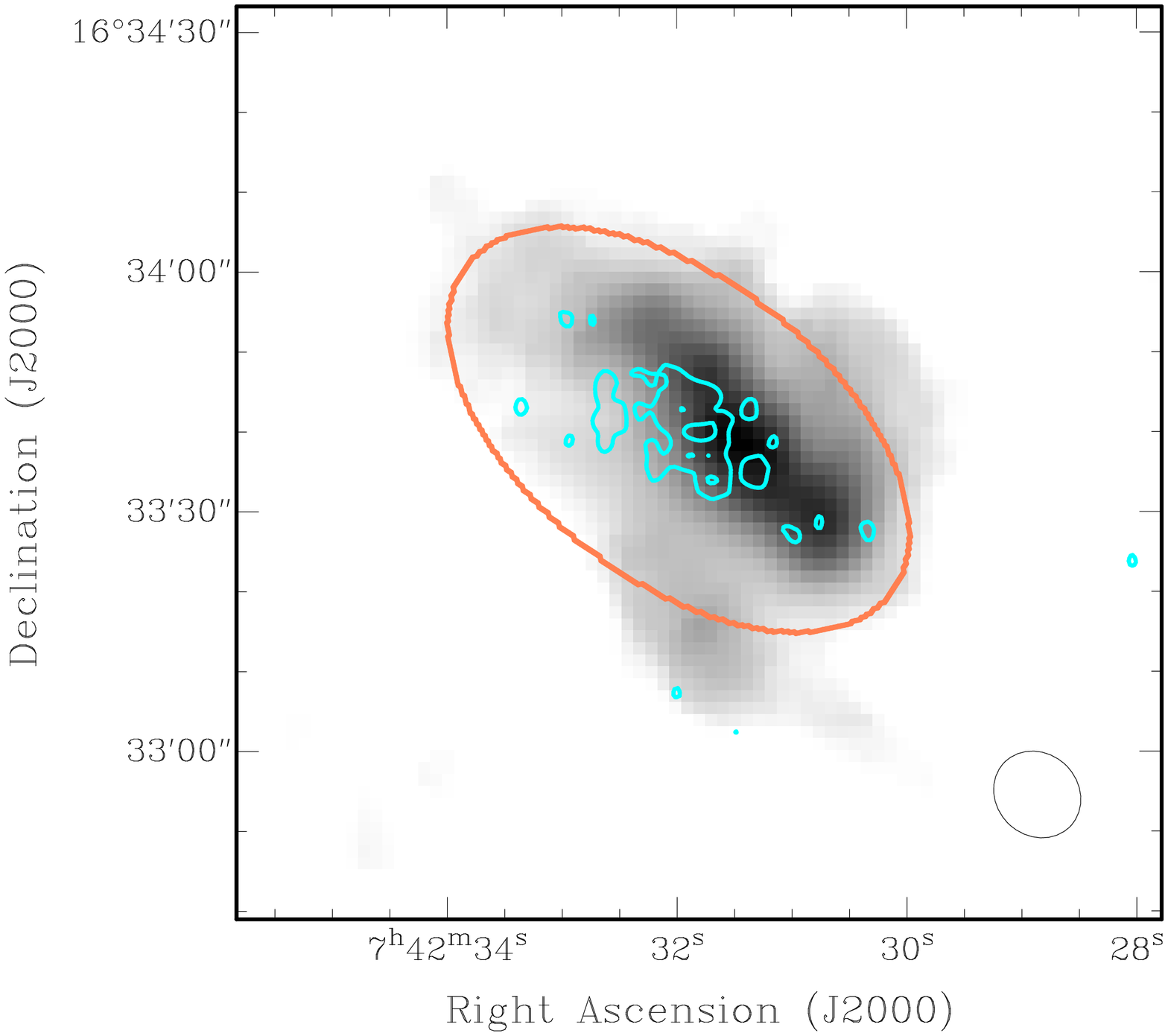}}}&
{\mbox{\includegraphics[height=4cm,angle=270]{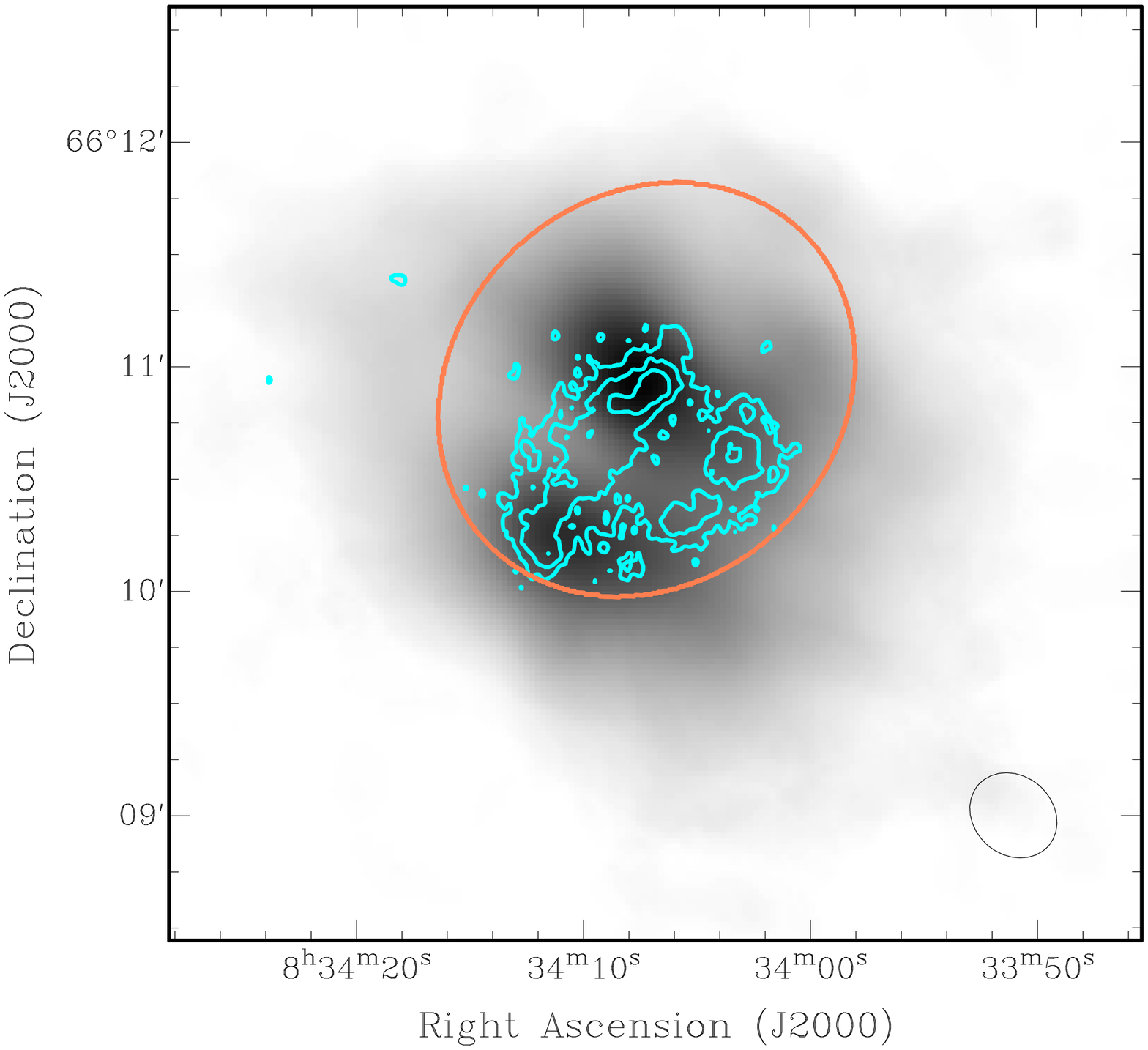}}}\\
\\
UGC 685&UGC 3755&KK 65&UGC 4459\\
{\mbox{\includegraphics[height=4cm,angle=270]{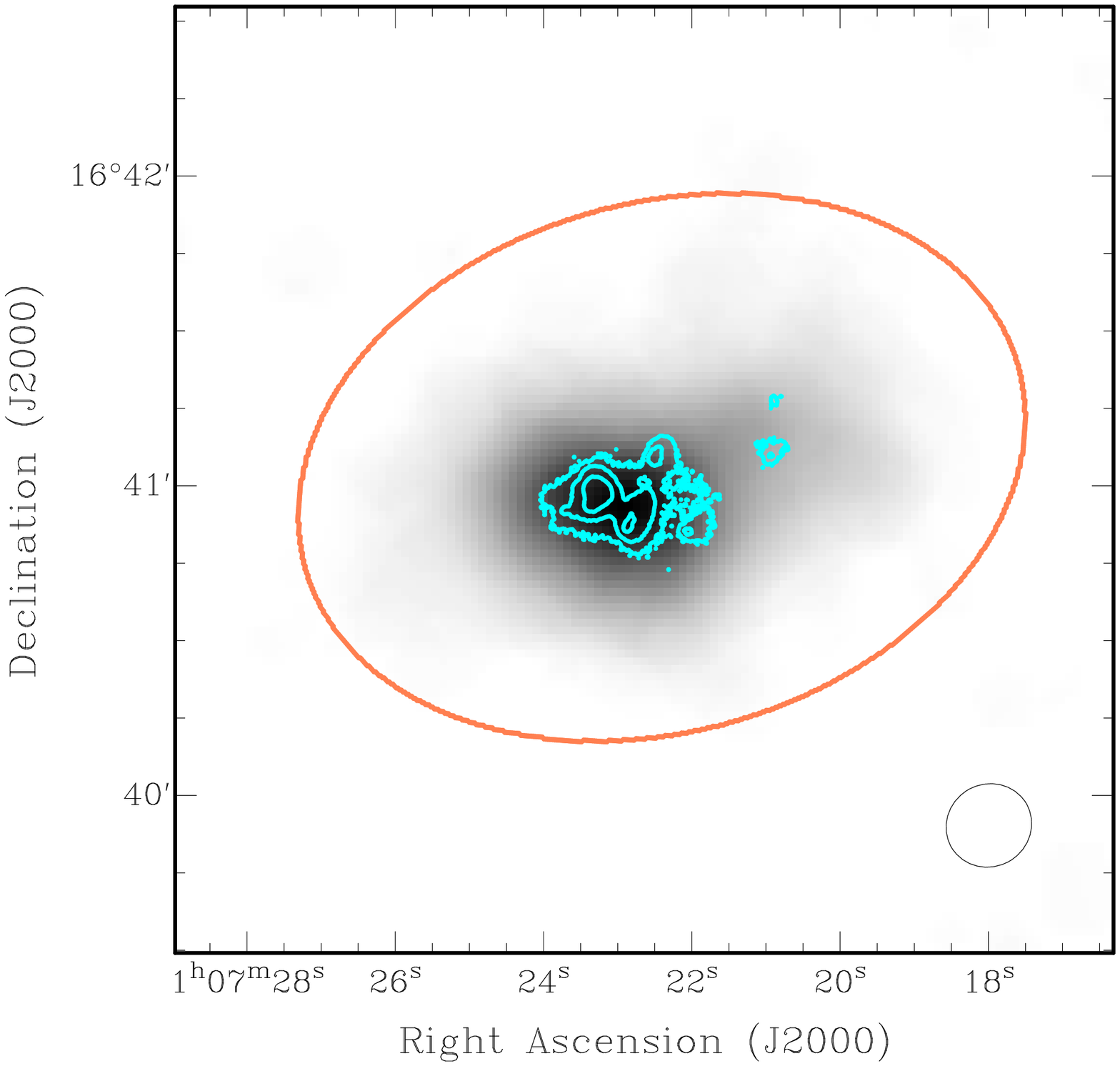}}}&
{\mbox{\includegraphics[height=4cm,angle=270]{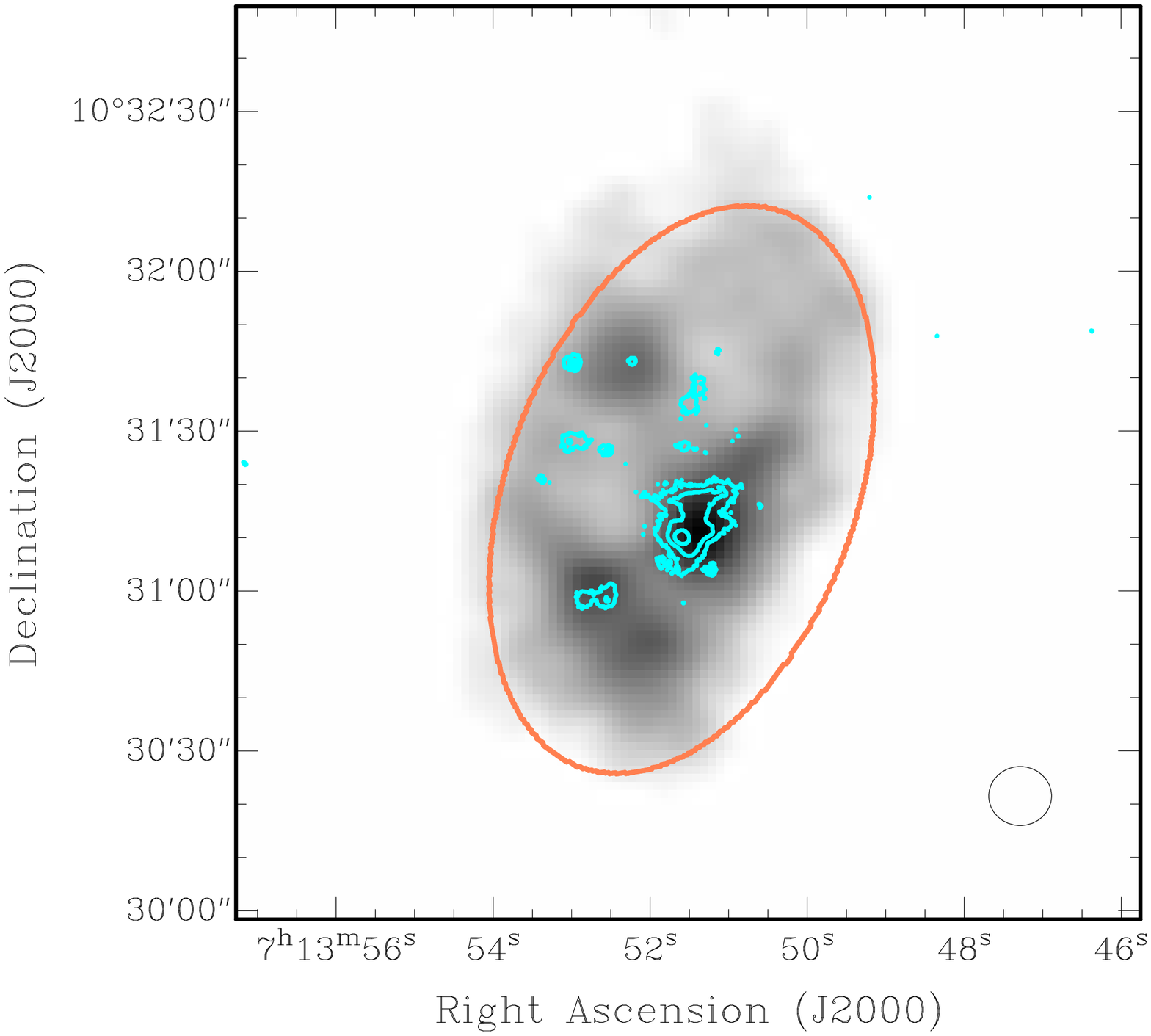}}}&
{\mbox{\includegraphics[height=4cm,angle=270]{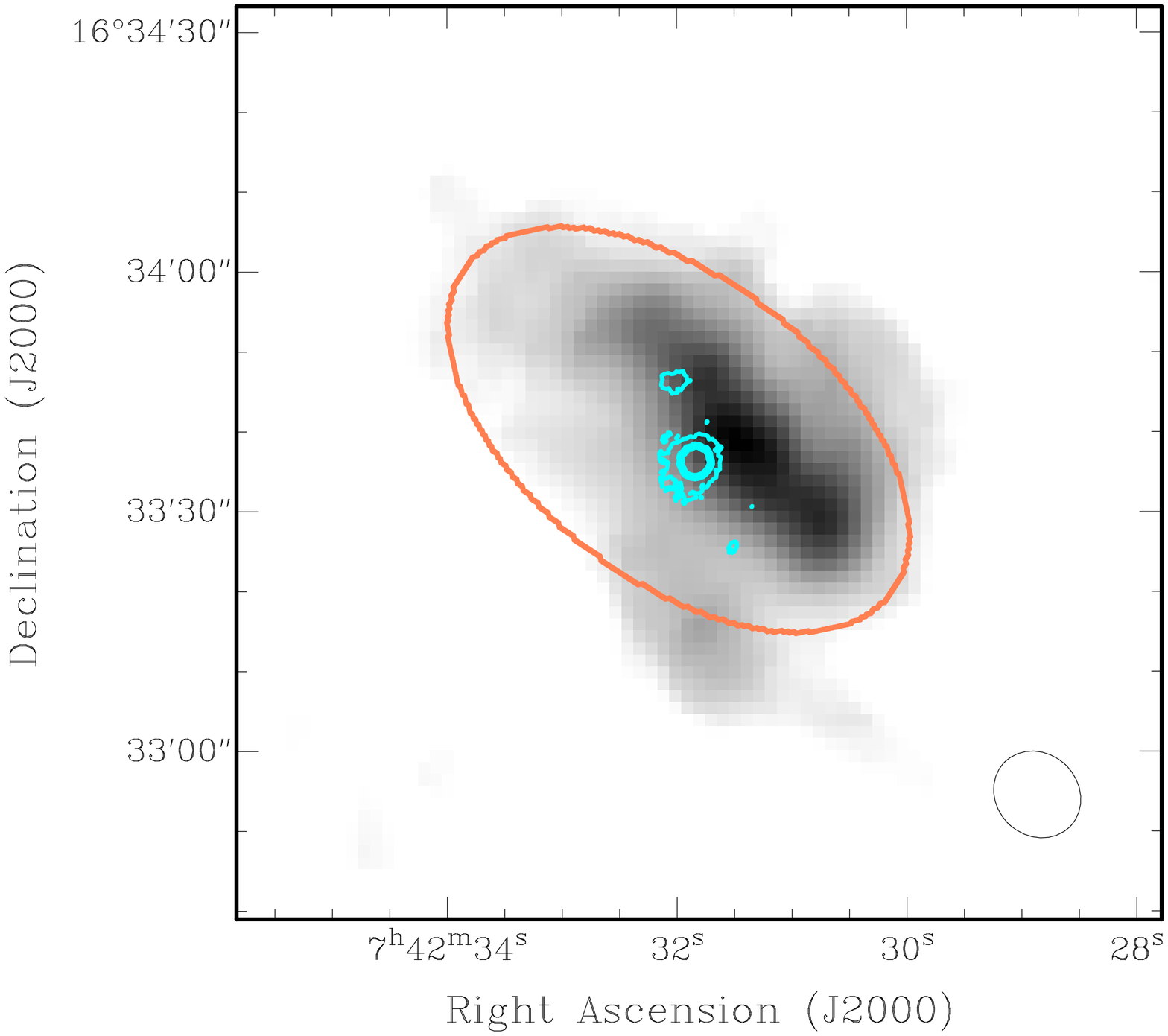}}}&
{\mbox{\includegraphics[height=4cm,angle=270]{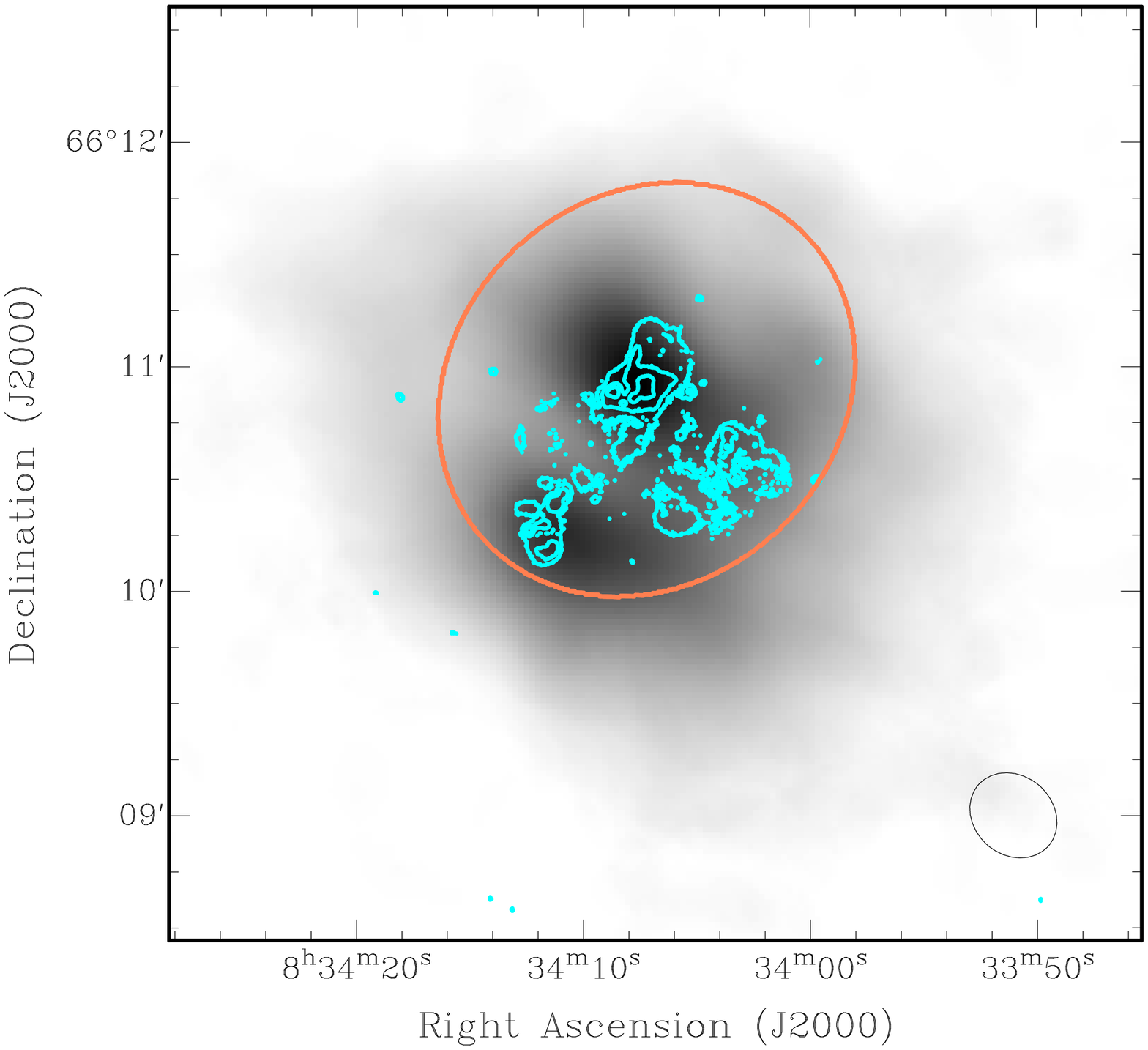}}}\\
\\
DDO 6&KK 41&DDO43&UGCA 438\\
{\mbox{\includegraphics[height=4cm,angle=270]{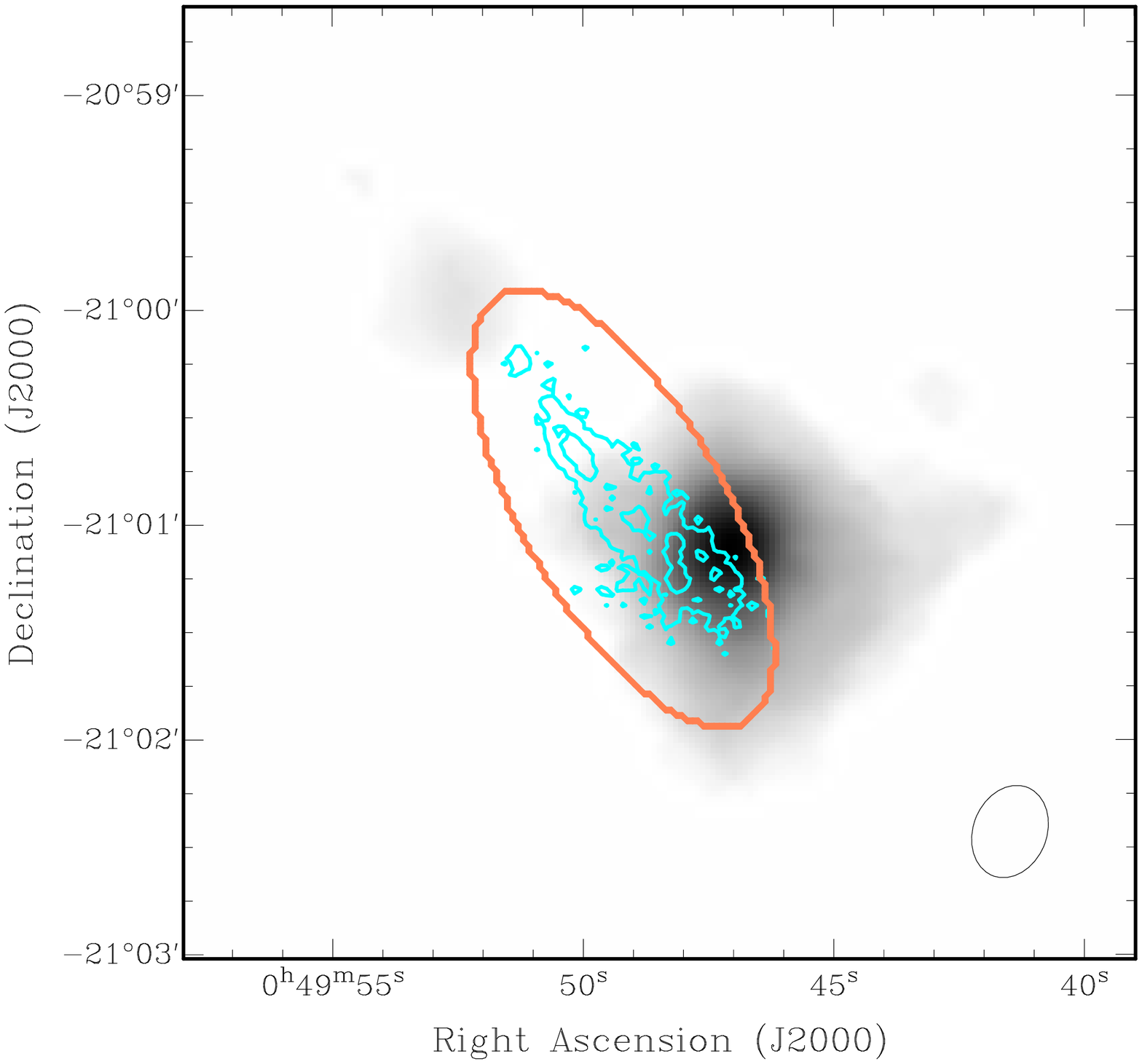}}}&
{\mbox{\includegraphics[height=4cm,angle=270]{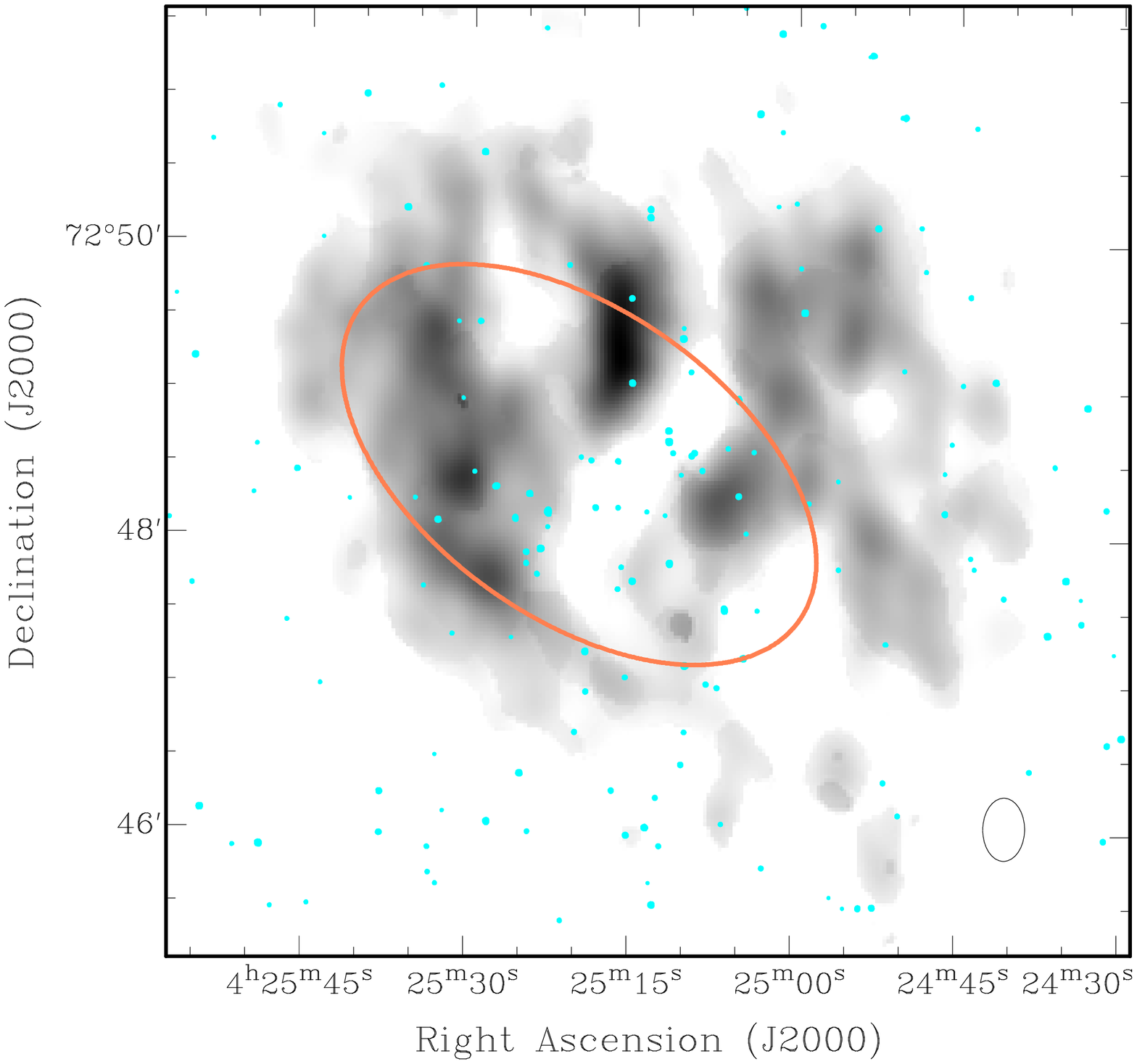}}}&
{\mbox{\includegraphics[height=4cm,angle=270]{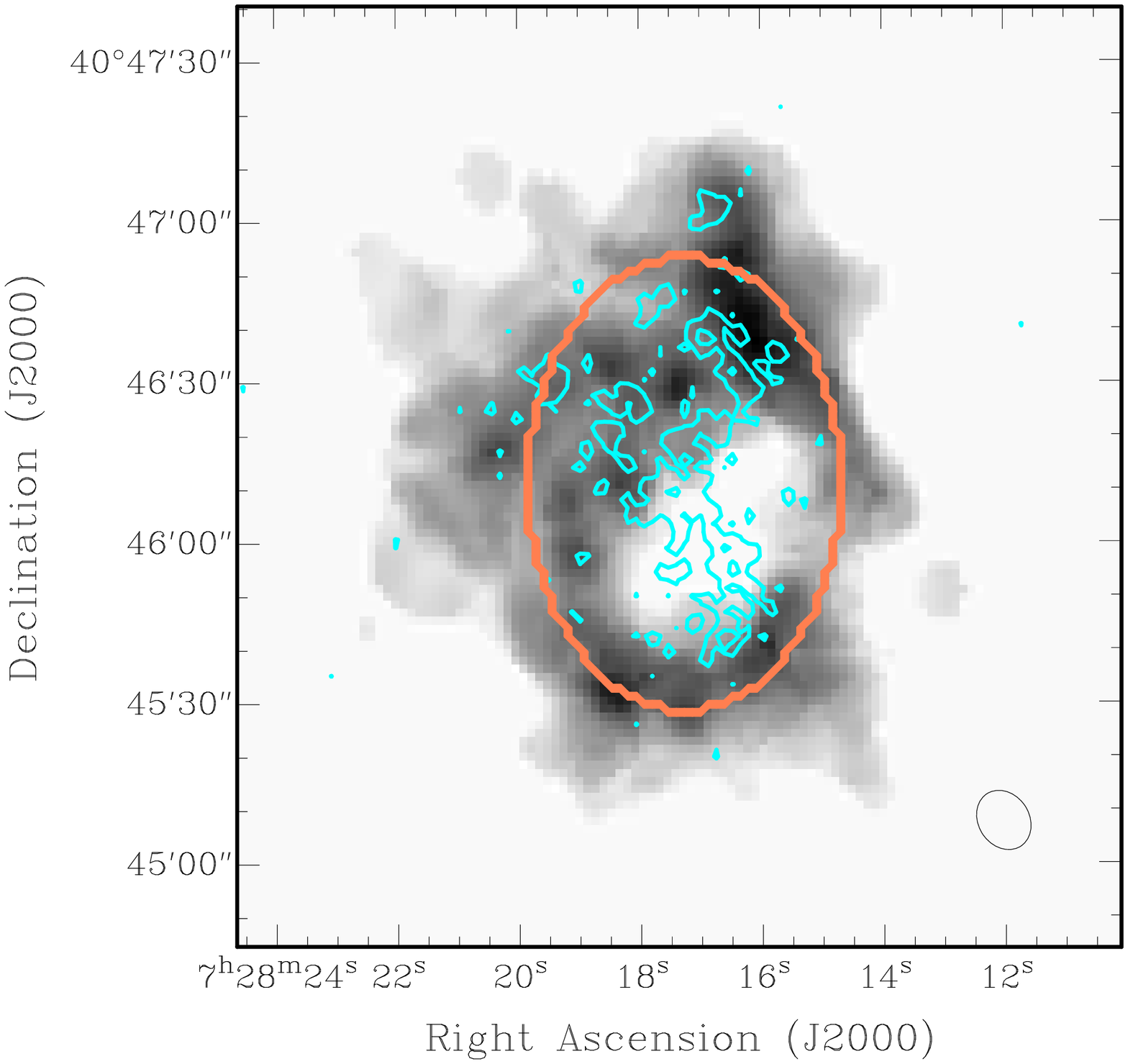}}}&
{\mbox{\includegraphics[height=4cm,angle=270]{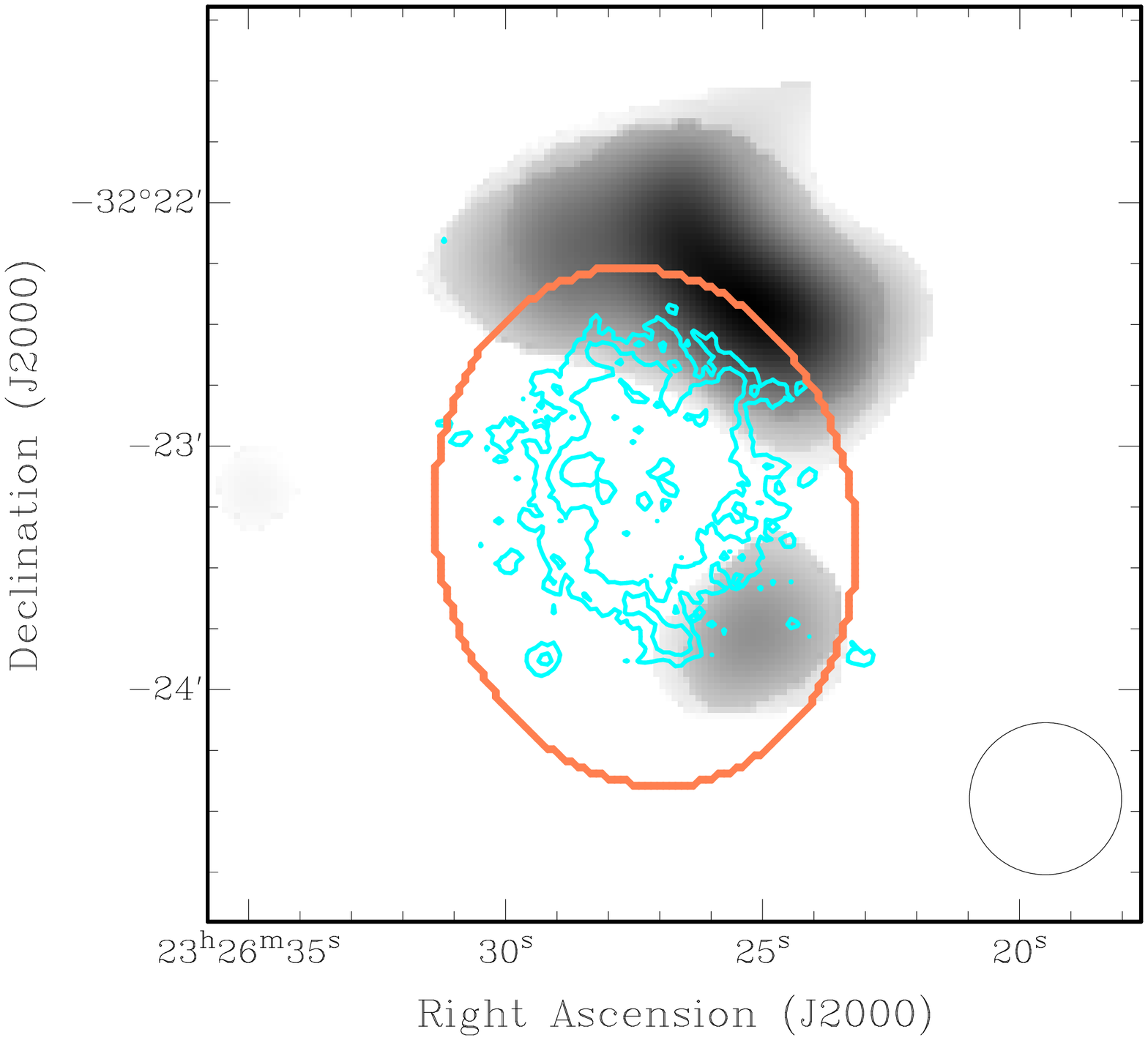}}}\\
\end{tabular}
\caption{SFR indicator contours in cyan overlayed on HI column density in greyscale for representative sample galaxies (for the full set of overlays see the online version). The upper and middle rows show a set of galaxies with undisturbed HI morphology, with  FUV emission in cyan for the upper row and \ha\ emission in cyan for the middle row. The lower row shows a set of galaxies with disturbed HI morphology with FUV emission in cyan. The Holmberg ellipse in each case is shown in orange. The level of the first contour is arbitrarily chosen so that traces of background emission are present, and subsequent contours are in multiples of 4. The HI beams are shown at the bottom right corner of each panel.}
\label{fig:ov}
\end{center}
\end{figure*}

\bsp

\label{lastpage}


\begin{thebibliography}{}

\bibitem[\protect\citeauthoryear{Asplund et al.}{2009}]{asp09} Asplund M., Grevesse N., Jacques Sauval A., Scott P., 2009, ARA\&A, 47, 481
\bibitem[\protect\citeauthoryear{Begum et al.}{2008}]{beg08} Begum A., Chengalur J.~N., Karachentsev I.~D., Sharina M.~E., Kaisin S.~S., 2008, MNRAS, 386, 1667
\bibitem[\protect\citeauthoryear{Bigiel et al.}{2010}]{big10} Bigiel F., Leroy A., Walter F., Blitz L., Brinks E., de Blok W.~J.~G., Madore B., 2010, AJ, 140, 1194
\bibitem[\protect\citeauthoryear{Bolatto, Wolfire \& Leroy}{2013}]{bol13} Bolatto A. D., Wolfire M., Leroy A. K., 2013, ARA\&A, 51, 207
\bibitem[\protect\citeauthoryear{Bolatto et al.}{2011}]{bol11} Bolatto A. D. et al., 2011, ApJ, 741, 12
\bibitem[\protect\citeauthoryear{Bouchard, Da Costa \& Jerjen}{2009}]{bou09} Bouchard A., Da Costa G. S., Jerjen H., 2009, 137, 3038
\bibitem[\protect\citeauthoryear{Buyle et al.}{2006}]{buy06} Buyle P., Michielsen D., De Rijcke S., Ott J., Dejonghe H., 2006, MNRAS, 373, 793
\bibitem[\protect\citeauthoryear{Carraro et al.}{2007}]{car07} Carraro G., Geisler D., Villanova S., Frinchaboy P.~M., Majewski S.~R., 2007, A\&A, 476, 217 
\bibitem[\protect\citeauthoryear{Dale et al.}{2009}]{dal09} Dale D. A. et al., 2009, ApJ, 703, 517
\bibitem[\protect\citeauthoryear{da Silva, Fumagalli \& Krumholz}{2014}]{daS14} da Silva R. L., Fumagalli M., Krumholz M. R., 2014, arXiv:1403.4605
\bibitem[\protect\citeauthoryear{Doyle \& Drinkwater}{2006}]{doy06} Doyle M. T., Drinkwater M. J., 2006, MNRAS, 372, 977
\bibitem[\protect\citeauthoryear{Ekta \& Chengalur}{2010}]{ekt10} Ekta B., Chengalur J. N., 2010, MNRAS, 406, 1238
\bibitem[\protect\citeauthoryear{Elmegreen et al.}{2013}]{elm13} Elmegreen B. G., Rubio M., Hunter D. A., Verdugo C., Brinks E., Schruba A., 2013, Nat., 495, 487
\bibitem[\protect\citeauthoryear{Engelbracht et al.}{2008}]{eng08} Engelbracht C. W., Rieke G. H., Gordon K. D., Smith J.-D. T., Werner M. W., Moustakas, J., Willmer C. N. A., Vanzi, L., 2008, 678, 804
\bibitem[\protect\citeauthoryear{Fu et al.}{2013}]{fu13} Fu J. et al., 2013, MNRAS, 434, 1531
\bibitem[\protect\citeauthoryear{Genovali et al.}{2014}]{gen14} Genovali K., et al., 2014, A\&A, 566, A37
\bibitem[\protect\citeauthoryear{Governato et al.}{2010}]{gov10} Governato F. et al., 2010, Nature, 463, 203
\bibitem[\protect\citeauthoryear{Hao et al.}{2011}]{hao11} Hao C.-N., Kennicutt Jr. R. C., Johnson B. D., Calzetti D., Dale D. A., Moustakas J., 2011, ApJ, 741, 124
\bibitem[\protect\citeauthoryear{Hermanowicz, Kennicutt \& Eldridge}{2013}]{her13} Hermanowicz M. T., Kennicutt R. C., Eldridge J. J., 2013, 432, 3097
\bibitem[\protect\citeauthoryear{Hopkins et al.}{2013}]{hop13} Hopkins P. F., Keres D., Onorbe J., Faucher-Giguere C.-A., Quataert E., Murray N., Bullock J. S., 2013, arXiv:1311.2073
\bibitem[\protect\citeauthoryear{Hunter, Elmegreen \& Ludka}{2010}]{hun10} Hunter D. A., Elmegreen B. G., Ludka B. C., 2010, AJ, 139, 447
\bibitem[\protect\citeauthoryear{Kaisin \& Karachentsev}{2008}]{kai08} Kaisin S. S., Karachentsev I. D., 2008, A\&A, 479, 603
\bibitem[\protect\citeauthoryear{Kaisin, Karachentsev \& Kaisina}{2011}]{kai11} Kaisin S. S., Karachentsev I. D., Kaisina E. I., 2011, Ap, 54, 315
\bibitem[\protect\citeauthoryear{Karachentsev \& Kaisin}{2007}]{kar07} Karachentsev I. D., Kaisin S. S., 2007, AJ, 133,1883
\bibitem[\protect\citeauthoryear{Karachentsev \& Kaisin}{2010}]{kar10} Karachentsev I. D., Kaisin S. S., 2010, AJ, 140, 1241
\bibitem[\protect\citeauthoryear{Karachentsev \& Kaisina}{2013}]{kar13a} Karachentsev I. D., Kaisina E. I., 2013, AJ, 146, 46
\bibitem[\protect\citeauthoryear{Karachentsev et al.}{2013}]{kar13} Karachentsev I. D. et al., AJ, 145, 101
\bibitem[\protect\citeauthoryear{Kauffmann}{2014}]{kau14} Kauffmann G., 2014, 441, 2717
\bibitem[\protect\citeauthoryear{Kennicutt}{1998}]{ken98} Kennicutt Jr. R. C., 1998, ApJ, 498, 541
\bibitem[\protect\citeauthoryear{Kennicutt \& Evans II}{2012}]{ken12} Kennicutt Jr. R. C., Evans II N. J., 2012, ARA\&A, 50, 531
\bibitem[\protect\citeauthoryear{Kennicutt et al.}{2008}]{ken08} Kennicutt Jr. R. C., Lee J. C., Funes J. G., Sakai S., Akiyama S., 2008, ApJS, 178, 247
\bibitem[\protect\citeauthoryear{Kennicutt et al.}{2009}]{ken09} Kennicutt Jr. R. C. et al., 2009, ApJ, 703, 1672
\bibitem[\protect\citeauthoryear{Krumholz}{2013}]{kru13} Krumholz M. R., 2013, MNRAS, 436, 2747
\bibitem[\protect\citeauthoryear{Lee et el.}{2009b}]{lee09} Lee J. C. et al., 2009, ApJ, 706, 599
\bibitem[\protect\citeauthoryear{Leitherer et al.}{1999}]{lei99} Leitherer C. et al., 1999, ApJSS, 123, 40
\bibitem[\protect\citeauthoryear{Leroy et al.}{2005}]{ler05} Leroy A., Bolatto A. D., SImon J. D., Blitz L., 2005, ApJ, 625, 763
\bibitem[\protect\citeauthoryear{Leroy et al.}{2012}]{ler12} Leroy A. K. et al., 2012, AJ, 144, 3
\bibitem[\protect\citeauthoryear{Leroy et al.}{2013}]{ler13} Leroy A. K. et al., 2013, AJ, 146, 19
\bibitem[\protect\citeauthoryear{Mac Low \& Ferrara}{1999}]{mac99} Mac Low M.-M., Ferrara A., 1999, ApJ, 513, 142
\bibitem[\protect\citeauthoryear{Marble et al.}{2010}]{mar10} Marble A. R. et al., 2010, ApJ, 715, 506
\bibitem[\protect\citeauthoryear{McKee \& Ostriker}{2007}]{McK07} McKee C. F., Ostriker E. C., 2007, ARA\&A, 45, 565
\bibitem[\protect\citeauthoryear{Meurer et al.}{2006}]{meu06} Meurer G. R. et al., 2006, ApJSS, 165, 307
\bibitem[\protect\citeauthoryear{Meurer et al.}{2009}]{meu09} Meurer G. R. et al., 2009, ApJ, 695, 765
\bibitem[\protect\citeauthoryear{Momose et al.}{2013}]{mom13} Momose R. et al., 2013, ApJ, 772, L13
\bibitem[\protect\citeauthoryear{Moustakas \& Kennicutt}{2006}]{mou06} Moustakas J., Kennicutt Jr. R. C., 2006, ApJS, 164, 81
\bibitem[\protect\citeauthoryear{Ostriker, McKee \& Leroy}{2010}]{ost10} Ostriker E> C., McKee C. F., Leroy A. K., 2010, ApJ, 721, 975
\bibitem[\protect\citeauthoryear{Pilyugin \& Thuan}{2005}]{pil05} Pilyugin L. S., Thuan T. X., 2005, ApJ, 631, 231
\bibitem[\protect\citeauthoryear{Raiter, Schaerer \& Fosbury}{2010}]{rai10} Raiter A., Schaerer D., Fosbury R. A. E., 2010, A\&A, 523, 64
\bibitem[\protect\citeauthoryear{Rela\~{n}o et al.}{2012}]{rel12} Rela\~{n}o M., Kennicutt Jr. R. C., Eldridge J. J., Lee J. C., Verley S., 2012, MNRAS, 423, 2933
\bibitem[\protect\citeauthoryear{Roychowdhury et al.}{2009}]{roy09} Roychowdhury S., Chengalur J. N., Begum A., Karachentsev I. D., 2009, MNRAS, 397, 1435
\bibitem[\protect\citeauthoryear{Roychowdhury et al.}{2011}]{roy11} Roychowdhury S., Chengalur J. N., Kaisin S. S., Begum A., Karachentsev I. D., 2011, MNRAS, 414, L55
\bibitem[\protect\citeauthoryear{Roychowdhury et al.}{2013}]{roy13} Roychowdhury S., Chengalur J. N., Karachentsev I. D., Kaisina E. I., 2013, MNRASL, 436, L104
\bibitem[\protect\citeauthoryear{Schaye \& Dalla Vecchia}{2008}]{sch08} Schaye J., Dalla Vecchia C., 2008, MNRAS, 383, 1210
\bibitem[\protect\citeauthoryear{Schruba et al.}{2011}]{sch11} Schruba A. et al., 2011, AJ, 142, 37
\bibitem[\protect\citeauthoryear{Schruba et al.}{2012}]{sch12} Schruba A. et al., 2012, AJ, 143, 138
\bibitem[\protect\citeauthoryear{Schmidt}{1959}]{s59} Schmidt M., 1959, ApJ, 129, 243
\bibitem[\protect\citeauthoryear{Silich \& Tenorio-Tagle}{2001}]{sil01} Silich S., Tenorio-Tagle G., 2001, ApJ, 552, 91
\bibitem[\protect\citeauthoryear{Taylor, Kobulnicky \& Skillman}{1998}]{tay98} Taylor C. L., Kobulnicky H. A., Skillman E. D., 1998, AJ, 116, 2746
\bibitem[\protect\citeauthoryear{Weidner \& Kroupa}{2005}]{wei05} Weidner C., Kroupa P., 2005, ApJ, 625, 754
\bibitem[\protect\citeauthoryear{Weisz et al.}{2012}]{wei12} Weisz D. R. et al., 2012, ApJ, 744, 44
\bibitem[\protect\citeauthoryear{Wyder et al.}{2009}]{wyd09} Wyder T. K. et al., 2009, ApJ, 696, 1834

\end{thebibliography}
\end{document}